\RequirePackage{fix-cm}
\documentclass[smallcondensed]{svjour3}     
\smartqed  
\usepackage{graphicx}
\usepackage{pbox}
\usepackage{hyperref}
\usepackage{booktabs} 
\usepackage{xspace}
\usepackage{multirow}
\usepackage{hhline}
\usepackage{fancybox}
\usepackage{balance}
\usepackage{epstopdf}
\usepackage{amssymb}

\usepackage{amsthm}
\usepackage{amsmath}
\usepackage{tikz}
\usetikzlibrary{positioning,shapes,fit,arrows}
\usepackage{xcolor}

\newcommand{\hypobox}[1]{

        \begin{center}\noindent\thicklines\setlength{\fboxsep}{8pt}\cornersize{0.2}\ovalbox{

                \begin{minipage}{3.0in}

                        \textit{#1}

                \end{minipage}} 

        \end{center}} 
\journalname{Empirical Software Engineering}
\doiname{Original Article in: https://doi.org/10.1007/s10664-019-09791-w}
\begin{document}

\title{Deriving a Usage-Independent Software Quality Metric }


\author{Tapajit Dey         \and
        Audris Mockus 
}


\institute{Tapajit Dey  \and Audris Mockus \at
            Department of Electrical Engineering and Computer Science \\  
            University of Tennessee, Knoxville \\
            Knoxville, Tennessee, USA \\
            \email{tdey2@vols.utk.edu}, \email{audris@utk.edu} \\
}

\date{}

\maketitle
\vspace{-50pt}

\begin{abstract}
Context: The extent of post-release use of software affects the number of
 faults, thus biasing quality metrics and adversely affecting
 associated decisions. The proprietary nature of usage data limited 
 deeper exploration of this subject in the past.
Objective: To determine how software faults and software use are
   related and how, based on that, an accurate quality 
   measure can be designed.  
Method: Via Google Analytics we measure new users, usage intensity,  usage frequency, 
  exceptions, and release date and duration for complex 
  proprietary mobile applications for Android and iOS.
  We utilize Bayesian Network and Random Forest 
  models to explain the interrelationships and to derive the usage
  independent release quality measure. To increase external validity, 
  we also investigate the interrelationship among various code complexity measures,
  usage (downloads), and number of issues for 520 NPM packages. We derived a 
  usage-independent quality measure from these analyses, and applied it on
  4430 popular NPM packages to construct timelines for comparing the perceived quality
  (number of issues) and our derived measure of quality during the lifetime of these 
  packages.
Results: We found the number of new users to be 
  the primary factor determining the number of exceptions, and found no
  direct link between the intensity and frequency of software usage and 
  software faults.  Crashes increased with the power of 1.02-1.04 of new user for the 
  Android app and power of 1.6 for the iOS app. 
  Release quality expressed as crashes per user was independent of
  other usage-related predictors, thus serving as a usage independent
  measure of software quality. Usage also affected quality in NPM, where
  downloads were strongly associated with numbers of issues, even after taking the other
  code complexity measures into consideration.
  Unlike in mobile case where exceptions per user decrease over time, for 
  45.8\% of the NPM packages the number of issues per download increase.
Conclusions: We expect our result and our proposed quality measure will help
  gauge release quality of a software more accurately and inspire further 
  research in this area.

\end{abstract}

\section{Introduction}\label{s:intro}

Improving quality of software is one of the objectives of software
engineering.  ``Software Quality'' has been defined in various ways,
however, in this paper we take a narrow focus on the manifestation of software
defects as crashes observed from the perspective of the
users. Observing a software crash, generally speaking, is a
manifestation of low quality of the software to a user. 
Thus, it seems intuitive to measure the quality of software\footnote{in fact, we
  mean to measure one aspect of the quality of software}
by counting the number of crashes, with more crashes being associated
with lower quality. If, for example, we compare 
two different softwares or two different releases of a software,
we first calculate the number of crashes for each and then compare
these numbers. However, software with more users tends
to see more crashes~\cite{dey2018modeling,hmps15,IQ08} as each user
may exercise it differently. In
an extreme case, a software or a release with no users will have
no crashes, regardless of its quality. This interdependence of
software usage volume and crashes experienced is typically not
considered in quality measurement in industry or in empirical studies (although few studies
do note that~\cite{fenton2008using,fenton1999critique}).
Ignoring this relationship, however, would misguide quality
improvement efforts (avoid quality improvements for releases/
software with low usage) and/or misguided developer
performance metrics (reward developers of low-usage products). 
This analogy can also be extended for software
defects (bugs), and by extension for issues raised against a
software, since software crashes are manifestations of underlying
defects and~\cite{caper,hmps15} observed that the number of
discovered software defects increases with the number of users,
although the relationship between crashes and defects is not very
well understood~\cite{fenton1999critique}.  

One possible reason for this oversight is the scarcity of reliable usage data. While the number of defects and crashes reported by users are carefully tracked by most large scale projects (\textit{e.g.} Mozilla Firefox, Ubuntu etc.), tracking the variables related to usage, \textit{e.g.} the number of users, intensity of usage etc. is almost impossible without a reliable monitoring system. Such a system is rarely used by open-source software and even many traditional software-as-a-product systems do not or can not have such capability. Moreover, even when such a dataset is available, it is almost always proprietary, so obtaining and sharing it, even for the software development teams in these proprietary projects, is difficult since the deployment is typically managed by a different team within the organization. Without such data, however, it becomes exceedingly difficult to interpret the quality of a software from the customer reported crashes/defects alone due to the interdependence of usage and crashes/defects~\cite{dey2018modeling,hmps15,IQ08}.
The overarching goal of this study is to advocate the necessity of taking the usage aspect into consideration while comparing the qualities of different releases/softwares and to illustrate one possible way of doing so.

We were able to obtain the usage data for some mobile applications developed by Avaya, \textit{viz.}  Avaya Communicator for Android (currently known as Avaya Equinox\textregistered) and Avaya one-X\textregistered  Mobile SIP for iOS. The usage data was obtained from Google Analytics. 
For analyzing the usage data for these applications, we used three usage related variables: number of users, 
usage intensity (average duration of software use per user), and usage frequency 
(average number of times the app was used by a user), along with two variables describing attributes 
of the particular release: release date and effective duration of the release, measured by how 
long the release continued to have new users, and looked at how these variables affect 
the number of exceptions \textit{i.e.} application crashes.  Thus, the \textbf{first research question} we are addressing in this paper is about modeling the relationships between these different post-deployment variables, specifically, finding the relationships among variables describing different aspects of software
usage and software crashes (which are manifestations of underlying defects). 

Since the quality of a software/release measured by the number of defects/crashes is often misleading due to its dependence on usage, as we mentioned before, our \textbf{second research question} is about constructing a usage-independent measure of quality, which would give us the ability to compare the quality of software releases more accurately.

The methodology we employed for addressing these research questions is as follows: After the usual data cleaning and variable construction stages, we  used a Bayesian Network (BN) model to discover the interrelationship between the variables. The BN model was generated by using structure search algorithms, and the search method was chosen based on the result of a simulation study. We have presented the detailed result of the simulation study and hope that other practitioners willing to use BN structure search methods in their work would find it useful.  Then, we ran a Random Forest (RF) regression model to identify the important variables for modeling the number of exceptions. All analyses in this study was done in R~\cite{R}. We found that the frequency and intensity of usage have little impact on the number of exceptions, but the number of users does have a significant impact. Thus, we establish the interdependence between the number of crashes and usage, specifically, the number of users, and finally propose a quality metric that is independent of usage, which would enable us to compare the qualities of different softwares and/or different releases of a software more accurately. 

In our previous work~\cite{dey2018modeling}, we only analyzed the
General Availability releases for  Avaya Communicator for
Android. Since all of the data was from Google Analytics, we had
the same set of variables for all the Avaya softwares. We found
similar results from that study as well, with the number of new
users being the most important factor affecting the number of
exceptions. 
Furthermore, we proposed a quality metric of average number of
exceptions experienced by end users (so lower means better) and
found it to be independent of other usage metrics.  

In this study, we have added the analysis of the development version 
of Avaya Communicator for Android and the
General Availability releases of Avaya one-X\textregistered  Mobile
SIP iOS Client. Although we collected data for several other apps,
those were dropped due to having too few releases/ exceptions/ users
to give a reliable result. We employed some new data correction steps 
for correcting the observed number of users and visits (Section~\ref{s:datapre}).
We used the same three modeling techniques and the same quality measure
 and found the result to be very similar for all cases considered. 
We also added a timeline showing how the perceived quality of
the releases vary with time for different releases.

To examine whether adding code complexity measures have any effect on our findings,
we also consider the relationship between downloads, a measure of usage, various code 
quality measures, specifically the average per-function count of logical
lines of code, cyclomatic complexity, Halstead effort, parameter count, and 
the average per-module maintainability index, and number of issues, a measure 
similar to the number of crashes or bugs, for 520  different Node Package Manager (NPM)
packages.  NPM is
the package manager for node.js, an open-source, cross-platform
JavaScript run-time environment.  NPM packages are commonly used for web
applications, including mobile applications, thus the software may, at least in some ways, be comparable to the Avaya Communicator.    Since we do not have the number of
crashes for these packages, we looked at the number of issues
reported for these packages instead. The code complexity measures for these packages
were obtained using the npm-miner dataset~\cite{chatzidimitriou2018npm}.
We selected 520 out of the 2000 packages in the dataset based on the criteria that 
the release for which the complexity measures are reported should be more 
than a month older when the the measures are calculated, since we 
looked at the total downloads and issues over the period of one month for these packages.
Our \textbf{third research question} is about investigating whether the usage (downloads) 
has any significant impact on the number of issues even after taking the code 
complexity measures into account?

Our \textbf{fourth and final research question} is about examining how the basic quality 
measure, the number of crashes for the mobile applications, and the number of issues
for the NPM packages vary with time and how it compares to our derived quality 
measure. We constructed timeline plots for answering this question, and looked at the
different releases of the mobile applications. For the NPM packages, we decided to 
broaden our scope, and  look at 4430 NPM packages
that had more than 10,000 monthly downloads and a GitHub page with issues enabled.
We found that for only 36 out of 4430 packages (0.8\%) the number of daily 
downloads is not a significant predictor (p-value $> 0.5$ ) of the number of
issues on that day, and constructed timeline plots comparing the two trends
(number of issues and downloads) for these packages. 

The primary contributions based on our results include an observed strong
relationship between the number of exceptions (crashes) and usage by
analyzing two different mobile softwares (three different versions)
and a quality metric that gives a more actionable measure of 
quality that tries to separate factors that are beyond the control
of the development team. The analysis of the NPM packages showed that more usage
  is associated with the increase in the number of issues beyond
  what can be explained by code complexity metrics. The wider implications of
these findings suggest the potentially serious problems in existing
quality metrics and predictions. Specifically, the organizational
goals for a software project quality should take into account usage,
and software defect predictors could be improved substantially if
the population of users would be taken into account and could be
predicted. Moreover, the analysis of the NPM packages established the
extendibility of the the concept, thus opening the possibility of
wider application of the approach. We have also presented the 
detailed result of the simulation study
we conducted to choose the best performing BN structure search
algorithms, which we believe will of of use to practitioners willing
to use BN structure search methods in their work. Finally, we
have presented a timeline showing how the perceived quality changes
with time during the different releases for the mobile applications
and during the life of the NPM packages, and discussed the trends we
found.

The rest of the paper is organized as follows: We discuss the research questions and the motivation for the study in Section~\ref{s:motiv}. In Section~\ref{s:data}, we describe the data used in our study, providing details of the data source, data collection process, and detailed data preprocessing steps.
In section~\ref{s:method}, we provide a overview of the methodology we used in the study, which includes details of the simulation study we performed for choosing the best performing BN structure search method that was used in subsequent analysis. In Section~\ref{s:result}, we present the answers to the research questions of our study, which includes the models describing the interrelationship between exceptions and other post-deployment variables, the quality measure we proposed and models showing its independence of other usage measures, and the results of the analysis of the NPM packages. We discuss various aspects of our result in Section~\ref{s:implication}. In Section~\ref{s:relwork}, we discuss various works in related topics. Finally, we discuss the limitations of our study in Section~\ref{s:limitation} and conclude the paper in Section~\ref{s:conclusion}.  

\vspace{-10pt}
\section{Motivation}\label{s:motiv}

We hypothesize that the observed number of software failures, measured by the number of defects, crashes, and/or issues reported against it, depend on a number of internal as well as external factors. Mathematically,
$ OSF = f(v_i, v_e) $, where OSF = observed software faults, $f(.)$ is some function, $v_i$ is a
 set of internal factors, and $v_e$ is a set of external factors. The internal factors represent
 the set of factors that are artifacts of the software development process, and includes product
 parameters like size of the software and code complexity, process parameters like change 
entropy (distribution of changes with a component), human factors like the number of authors
 involved, number of reviewers with expertise, and code review parameters like number of reviews,
 number of reviewers etc. The impact of of these parameters are well known and have been used 
in multiple past studies~\cite{mcintosh2015emse,mcintosh2014impact,rigby2013convergent,kononenko2015investigating}.
However, the number of \textbf{observed} software failures also 
depend on external factors that are not part of the software development process. 
Hypothetically, such parameters include the number of users using the product, the intensity of 
usage (for how long they use it), and frequency of usage (how frequently they use it). The 
effect of these parameters is not very known or studied. However, when the measures related to 
software failure are collected, these are always the \textbf{observed} measures, not the 
\textbf{actual} measures. It is impossible to know the \textbf{actual} number of bugs in a 
given software for instance. 

However, as we mentioned earlier, there are serious problems related to using the observed measures of failures for comparing the qualities of different softwares and releases, since the observed number of low defects/crashes could be a result of low usage, and not better quality. Therefore, to be able to compare the actual quality of different softwares/releases that is a function of only the internal factors, we need a measure that is free from the influence of external factors like usage, i.e. for such a quality measure Q, $Q = g(v_i)$, where $g(.)$ is some function.

It is worth mentioning that, hypothetically, the internal and external factors should be independent of each other, at least for the general availability releases of closed-source softwares, since most of the users of such releases are not involved in the software development process, thus the external usage factors should be unaffected by software design artifacts. Therefore, none of the internal or external factors should act as a confounding variable, since a confounder affects both the dependent and independent variables in a causal relationship. This means that not accounting for the external factors would not systematically misclassify the perceived quality of all releases/softwares under consideration, but would randomly misclassify some of them. Such random effects are much harder to detect, and could be one of the reasons why this topic hasn't gained as much attention, since such misclassification errors could have wrongly been attributed to measurement errors or just random effects. However, accounting for the effect of usage would have enabled the developers to correctly compare the qualities of these releases/softwares and explain many of these apparent random misclassification errors.

The (hypothetical) independence of external and internal factors
have implications on the design of this study as well. Since we are
using closed-source commercial mobile applications where the
development team did not explicitly track which code changes were
introduced for which release, we
could not precisely link source
code complexity measures to crash reports from each release. However, if we can establish the
interdependence between the usage (external) factors  and the
observed number of crashes, the validity of such relationship is
less likely to be affected by the presence of unobserved internal
factors. We, indeed, observe that to be the case for a large
  collection of NPM packages. Furthermore, the main point of adjusting the measure
  of quality for the
  external factors that are beyond the control of the development
  team, is to focus on the impact of internal factors that are,
  hopefully, under control of the development team.

Our study of the interrelationship between code complexity measures, extent of usage, and the
 number of issues show that (a) increased usage is associated
 with increase in the number of issues beyond what can be explained by
the code complexity measures, and (b) the BN model shows that usage 
(downloads) is independent of the code complexity measures. As noted
in the previous paragraph, this supports the results obtained in the
studies of the mobile applications, and suggests
the importance of control for the extent of usage before using the number of 
observed software faults as an internal measure of development quality.

\vspace{-20pt}
\section{Data Description}\label{s:data}
For this study, we looked into two different types of software, which are vastly different in nature. The primary focus of the study is on the commercial software developed by Avaya for mobile applications, which are from the telecommunication domain.  We chose the Avaya mobile applications because we had access to the actual post-deployment measures for these. We hypothesized that the number of users is the most important measure of usage, but, as we mentioned earlier, obtaining the actual number of users of a software post-deployment is extremely difficult without a dedicated monitoring tool, and such data is often proprietary. 
Therefore, having access to the actual usage measures gave us a golden opportunity to study the relationship between software usage and software crashes. However, using this data also had some limitations, e.g. we could not choose the variables being measured, the duration of the measurements, or the applications for which the data is being collected. Being a commercial software, the source code is essentially closed-source, which makes it difficult for us to conduct a through investigation with all variables of interest in one model.

For external validation of the theory developed using this data, 
and examining if the extent of usage has any effect on the number of observed software 
faults even after taking the various code complexity measures into account, we looked into the 
NPM packages, which are open-source JavaScript packages used in web-development. We used NPM 
for our study because (1) it is one of the largest open-source communities, which makes it a 
good candidate to be investigated, and (2) it collects the number of downloads for the packages,
 which is a far better measure than other usage measures like the number of stars or forks of 
GitHub projects. 

In this section, we provide some details about the software being studied, discuss the data source, describe the data, and give details about the data preprocessing steps.

\vspace{-10pt}
\subsection{Data on Mobile Applications developed by Avaya}

\vspace{-10pt}
\subsubsection{Software description}\label{sub:soft} 
One of the software chosen for this study was Avaya Communicator for
Android (currently known as Avaya Equinox\textregistered ). It
integrates the mobile devices of the users with their office Avaya
Aura\textregistered communications environment and delivers mobile
voice and video VoIP calling, cellular call integration, rich
conferencing, instant messaging, presence, visual voicemail,
corporate directory access and enterprise call logs\footnote{\url{https://support.avaya.com/products/P1574/avaya-equinox-for-android}}.

Another software we studied was the Avaya one-X\textregistered Mobile  SIP for  iOS, which provides mobile communications for the iPhone, iPod touch, and iPad through a wireless-enabled SIP Avaya Aura\textregistered environment combining enterprise features with the convenience of a mobile endpoint for users on the go. The Avaya one-X Mobile\textregistered SIP for iOS appears as an end point in the Aura\textregistered environment\footnote{\url{https://support.avaya.com/products/P0949/avaya-onex-mobile-sip-for-ios}}.

Avaya is developing large, complex, real-time software systems that
are embedded and standalone products. Development and testing are
spread through 10 to 13 time zones in the North America, USA, Europe
and Asia. R\&D department employed many virtual collaboration tools
such as JIRA, Git, WIKIs and Crucible. Development teams use
Scrum-like development methodologies with a typical 4-week
sprint. We consider a 15+ year old software component, the so-called
Spark engine.  As a software platform, Spark provides a consistent
set of signaling platform functionalities to a variety of Avaya
telephone product applications, including those of third parties.
Spark is a client platform that provides signaling manager, session
manager, media manager, audio manager, and video manager. The
codebase involves more than 200K files and, over all forks, over 4M
commits.  The Android software chosen for this study is a fork of
the Spark codebase. A more in-depth description of the development
process is provided in~\cite{amhp14}.

\vspace{-10pt}
\subsubsection{Data Description: Source}
The post-deployment data for the mobile applications were obtained from the 
Google Analytics platform.
Google Analytics is a web analytics service offered by
Google that tracks and reports website traffic. It is now one of the most
widely used web analytics services on the internet. In addition to
traditional web applications it also allows tracking of mobile
applications. To do that, the producer of a mobile application needs
to set up an account and instrument their mobile application to send certain
events to Google Analytics. Notably, it works for the mobile
applications investigated in this study. 

We collected data for a number of mobile applications developed by Avaya from Google Analytics, but some of the datasets turned out to be unusable for this study, for reasons ranging from very low volume of collected data (\textit{e.g.} Avaya Communicator for Android - Experimental Releases) to zero recorded exceptions making an analysis impractical (\textit{e.g.} Avaya One-X\textregistered   ScsCommander ). The following datasets were found usable:
\vspace{-10pt}
\begin{itemize}
    \item Avaya Communicator for Android - General Availability and Development versions.
    \item Avaya one-X\textregistered Mobile  SIP for  iOS - General Availability versions.
\end{itemize}
\vspace{-10pt}

The data was collected between December 2013 and May 2016, although the exact time varies across the applications. Although we are primarily focused on the General Availability (GA) versions, since only these versions are available for end-users, we also decided to look into the development version for Avaya Communicator for Android, since we have detailed data available for these versions and we wanted to see if it shows a different characteristics from the GA versions.  

The original data obtained from Google Analytics had measures for the variables 
listed in Table~\ref{t:measures}, aggregated at a per-day granularity, meaning 
that each entry in the original data table contained the measures for the numerical 
variables (marked with a $\dagger$ symbol in the table) for each unique combination
of date, application release version, operating system version, mobile device brand, 
category, and model. We had the same set of variable for all the applications listed above.
As we mentioned earlier, we had no role in selecting which variables to measure, and we received the data "as-is".

\emph{It is important to note that Google Analytics releases only
aggregate data even to developers of the application and limits the
number of REST API calls, so one can not, for example, retrieve
usage data for every calendar second or get exact time of the
events.} The daily counts split by release version of
the application, OS version, and type of device, provided
sufficiently fine granularity for our analysis. 
\begin{table}
\caption{Measures available in the Original Data}\label{t:measures}
\begin{tabular}{|p{6.5cm}|p{4cm}|}\hline
Application Release Version & No. of exceptions$\dagger$\\\hline
Operating System version in the user's device &Date of record entry\\\hline
No. of fatal exceptions$\dagger$& No. of new visits$\dagger$\\\hline
No. of visits$\dagger$ & Time on site$\dagger$\\\hline
\pbox{6cm}{Details on user's mobile device:\\ brand, category(mobile or tablet) and model} & No. of new users$\dagger$\\\hline
No. of total users$\dagger$& Sessions per user$\dagger$\\ \hline
\end{tabular}
\vspace{-10pt}
\end{table}

\vspace{-10pt}
\subsubsection{Data Preprocessing}\label{s:datapre}

This section contains the data cleaning, transformation, and variable 
construction steps undertaken prior to the application of the different modeling 
methods. The major preprocessing step is aggregating the measures to a per-release 
granularity. We had two main reasons for aggregating the data:
\begin{enumerate}
\item The goals of our study are concerned with identifying the relationship between exceptions and other post-deployment variables for different releases, and defining a quality measure to compare the qualities of different releases. Therefore, having the measures aggregated at per-release granularity is essential.
\item We would have been able to take a time-series based approach and still work out our goals if the releases were cleanly separated in time, i.e. if there were no overlap between releases. Unfortunately, we observed from the data that users continue to use one release long after the subsequent releases are available, and there is no clear pattern about how long a release is used. Therefore, we had to aggregate the data to a per-release level to be able to achieve the goals of this study.
\end{enumerate}

The preprocessing steps we took are discussed below:

\textbf{Removal of variables before aggregation: } Upon initial
investigation into the data, we found that no. of exceptions and
no. of fatal exceptions were exactly the same, as recorded by Google
Analytics, so we removed the no. of fatal exceptions from the
dataset. Only fatal exceptions were recorded for this application,
i.e., crashes that require a complete restart of the mobile
application and, potentially, may affect the operating system
itself.  This is not surprising since the bulk of the functionality
for the application was written in C$++$ and called from Android
Java applications via Native Interface.
We did not consider the variables related to mobile device
details and Android operating system versions because the
application, as noted above, was primarily written in C$++$ and the
user interface aspects that vary greatest among devices and versions
of OS were not likely to have influence. To validate that
assumption we investigated and found no
correlation of exceptions with either variable. 

\noindent
\textbf{Data correction:}
This additional prepossessing step was not a part of our previous study~\cite{dey2018modeling}.
We found during careful inspection of the data that some of the releases had non-zero number
of users but zero new users in the dataset. This obviously hints at some part of the data 
being missing. So, as a data correction step, we modified the number of new users so that 
in the chronological order of the data, the cumulative number of new users is never 
less than the number of users for a day.

\noindent
\textbf{Aggregating data to per-release granularity:}
We had some missing values in the data, however, most of the missing data was about the
mobile devices and since we didn't use them in our analysis, we got rid of that
problem by simply dropping the variables. Since our aim is to model the quality of the
different releases, we aggregated the data to a per-release 
granularity, from the the original data that was recorded in per-day granularity. 
The raw data contained 177 different GA releases and 25 development releases for the 
Avaya communicator for Android and 11 GA releases for the Avaya mobile SIP for iOS. 
We dropped 4 GA releases for the Avaya communicator for Android from further 
consideration because a significant
portion of observations were missing. The result of aggregation, however,
was two new variables: start date (first day for which we have a
record for that release) of a release, and end date (last date for
which we have a record for that release) of a release, which in turn
helped create another variable: duration of a release. We did
not to keep the end date in the final table, since duration and
start date can be used to compute the end date.

\noindent
\textbf{Verifying the correctness of Release date:}
The original data involves only the usage aspects and the version
information of the software. The project under consideration was
relatively new and it was one of the early attempts for the team to deploy
mobile software on Android and iOS. As such, not everything
was well documented and also was rapidly evolving over time  and no
record of the exact release dates for most of the releases was
available. We did manage to get release dates for some of releases
from Google Play Store/ Apple App Store, but not all the release dates  were
available.  For the releases with dates available on Google Play
Store/ Apple App Store, the official release dates from Avaya records, and the start
dates obtained from the data were either very close or exactly the
same, so we do not have a reason to doubt the dates obtained from the
data. 

\noindent
\textbf{Removal  of variables post aggregation:}
The numerical variables were aggregated to give a sum for each
variable. Upon further inspection, we found the number of users, new
users, visits, and new visits to be highly correlated. In the second
iteration, we removed the variable ``sessions per user'', because
aggregating it directly is meaningless, and we were not sure how it
was originally calculated by Google Analytics (was it a mean or
a median? were new users or total users counted?). We also removed
the ``total users'' and ``total visits'', because while summing up the
new users/visits for each day gives an accurate measurement of the total
number of new users/visits for a release, it is not guaranteed that
summing up total users/visits does the same due to possible double
counting the number of users/visits. 

\noindent
\textbf{Final list of variables:}
Keeping the goal of our study in mind, the variables we have after the initial cleaning 
steps give us necessary information for a model of post-release defects and software usage.
In our list of variables, we have the total number of exceptions \textit{i.e.} post-release defects. 
As for measures related to software usage, we have the total number of new users;
the ``Time.On.Site'' variable, normalized by the number of users of a release,
provides a measure for the temporal intensity of usage per user;
and the number of visits per user is a measure for the frequency of usage.
We also have two variables related to each individual release: the start date \textit{i.e.} the release
date gives a measure for the calendar time of each release, and is useful in gaining insight

about if the number of post-release defects and software usage vary with time, and 
the duration of a release, which could have an effect on the number of exceptions
and the number of new users, since these variables were not normalized with duration.
Since we only have a limited amount of data, we restricted ourselves to use only 
these six variables.
Our final aggregated data table had the measures listed in Table~\ref{t:finalvarss}, 
with the corresponding variable names we used in the model enclosed in brackets. 

\begin{table}
\caption{Measures in the Aggregated Data Table}\label{t:finalvarss}
\begin{tabular}{|l|l|}\hline
\pbox{6cm}{\textit{Release variable} - Start Date \\for the release (Release.Date)} & \pbox{6cm}{\textit{Release variable} - Effective Duration \\of the release (Release.Duration)}\\\hline
\pbox{6cm}{\textit{Post-Release defects} - \\Total No. of exceptions (Exceptions)}  &  \pbox{6cm}{\textit{Usage variable} - Average\\ time on site per user (Usage.Intensity)}\\\hline
\pbox{6cm}{\textit{Usage variable} -Total\\ number of new users (New.Users)} & \pbox{6cm}{\textit{Usage variable} \\- No. of visits per user (Usage.Frequency)}\\\hline
\end{tabular}
\vspace{-10pt}
\end{table}

To reiterate what we mentioned earlier, we had no control over which variables to measure during data collection, however, while the set of variables we obtained are not exhaustive, we believe the three usage related variables: number of new users, usage intensity, and usage frequency adequately capture and report how much usage a software is getting. 

\begin{figure}[!t]
\centering
\includegraphics[width=0.9\linewidth]{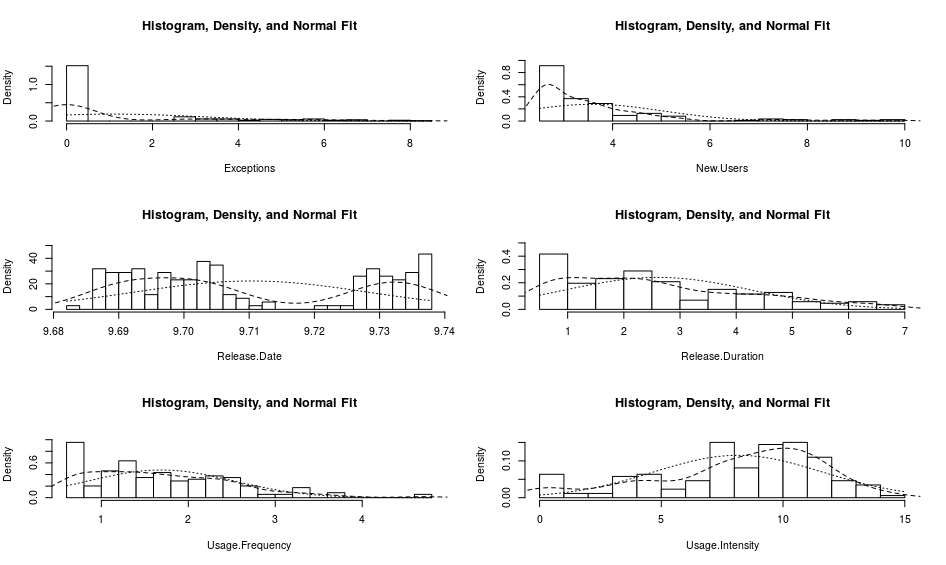}%
\caption{Distribution of the variables after transformation: GA releases of Avaya communicator for Android}
\label{fig:distr}
\vspace{-10pt}
\end{figure}

\noindent
\textbf{Log-transformation of variables:}
The release date was converted from the Date format to numeric format, which 
resulted in the values for the release date variable being represented by the 
difference in days from Unix time (counted from 1970-01-01).
We found that all of the variables under consideration
had a long-tailed distribution, so we took logarithm of them. 
The distribution of the variables of GA releases of Avaya communicator 
for Android is shown in Figure~\ref{fig:distr}.
The distribution of the variables of other applications  is available in our GitHub repository: \url{https://github.com/tapjdey/release\_qual\_model}.

\vspace{-10pt}
\subsection{The NPM Packages}

As mentioned before, we looked at 520 NPM packages for examining the interrelationship 
between the code complexity measures, the extent of usage, and the number of issues.
The code complexity measures for these packages were obtained from the npm-miner 
dataset~\cite{chatzidimitriou2018npm}, which contained information on 2000 packages 
(one particular release for almost all packages). However, we decided to only 
look at the package releases which were released more than a month before the data was 
collected, because we used the number of downloads over a month as our measure of usage (
since daily or even weekly download numbers tend to be quite noisy), and we ended up with 
520 releases of 520 different packages (one particular release per package).

However, for answering our fourth research question about exploring how our quality 
measure varies with time for the different NPM packages, and how it compares to the 
number of issues, the direct measure of observed software faults, we decided to broaden 
our scope to look at all NPM packages with  more than 10,000 downloads per month 
(according to~\cite{npmdl}, automated downloads are expected to be
around 50 per day, or 1500 per month, and packages with over 10K
downloads should, therefore, not be noticeably impacted by 
 downloads by automated sources.) and a GitHub page with issues.
 With this criteria we ended up with 4430 packages, which contained the 
520 packages we used for analysis earlier.

\vspace{-10pt}
\subsubsection{The NPM Ecosystem}

Node Package Manager or NPM is one of the most active and dynamic software ecosystems at present. It  hosted more than 800,000 packages at the time of data collection, and have more than doubled in size in past couple of years (in January 2017, NPM reportedly hosted around 350,000 packages~\cite{npmpkg}). The popularity of NPM packages have, accordingly, skyrocketed as well. According to~\cite{npmpop}, ``JavaScript is getting more popular all the time, and NPM is being adopted by an ever greater percentage of the JavaScript community." About 75\% of all JavaScript developers used NPM, with about 10 million users, in January 2018, according to~\cite{npmpop}. Therefore, NPM is an excellent candidate for this study. Moreover, since they track the number of downloads of all packages in the ecosystem, which, in spite of essentially being a mix of downloads by users, bots, and mirror servers, as explained in~\cite{npmdl}, is the closest measure of
usage we could find for open-source projects, and is a far better measure than, \emph{e.g.} number of stars of a GitHub repository which was used in studies like~\cite{borges2016understanding} as a measure of popularity, which is little different from usage as we measure it.

\vspace{-10pt}
\subsubsection{Measures collected from npm-miner dataset}
The npm-miner dataset~\cite{chatzidimitriou2018npm} contained information on 2000 NPM packages with  data from the following tools and APIs:    \textit{(1) eslint,  (2) escomplex, (3) nsp,  (4) eslint-security-plugin,   (5) jsinspect,     (6) sonarjs,   (7) npms.io,}  and   \textit{(8) GitHub}.
As mentioned before, we used 520 packages for our analysis in this study.
We used the monthly download numbers, collected from the analysis result of \textit{npms.io} 
and  the code complexity measures,  collected from the analysis result of 
\textit{escomplex}\footnote{https://www.npmjs.com/package/escomplex\#result-format}. 
In particular, we looked at the following code complexity measures for each NPM package, since they represent the average per function complexity measures for the packages:

\begin{itemize}
    \item \textit{loc}: The average per-function count of logical lines of code.
\item \textit{cyclomatic}: The average per-function cyclomatic complexity.
\item \textit{effort}: The average per-function Halstead effort.
\item \textit{params}: The average per-function parameter count.
\item \textit{maintainability}: The average per-module maintainability index.
\end{itemize}

\vspace{-10pt}
\subsubsection{NPM data: Defining collection parameters}
 We found 4430 projects which had more than 10,000
monthly downloads since January 2018 and also had public GitHub
repositories with nonzero number of issues.  We collected the number
of downloads and the total number of issues for all these packages
from 2015-03-01 to 2018-08-31. However, we did not conduct a release
by release comparison for these packages, because the release
durations vary by a lot for most packages. Since the recorded number
of downloads is a mix of downloads by human and non-human users, a
release by release comparison would not give a reliable picture of
the effect of actual usage by human users on the number of
issues. However, the number of downloads by bots are relatively
stable and vary only with time~\cite{npmdl}, so controlling for the date (time variable) would
eliminate the spurious effects of downloads by bots. So, we decided
to focus on the entire packages instead of releases of the packages,
and measured the effect daily downloads have on number of issues of
that package on that day after controlling for the calendar date.

\vspace{-10pt}
\subsubsection{NPM data: Data collection for the 4430 packages}

We used the API provided by NPM for collecting daily downloads of the 4430 NPM packages. 
(The API documentation is available in:\\ 
\textit{https://github.com/npm/registry/blob/master/docs/download-counts.md}).

To obtain the metadata information for every package in NPM, we wrote a ``follower" script, as described in 
\\ \textit{https://github.com/npm/registry/blob/master/docs/follower.md}.
The output contained the metadata information for all releases of all packages in NPM. 
From this we extracted the URL of GitHub repositories of the packages. Some NPM packages do not
 have a valid GitHub URL, so those were dropped from subsequent analysis, as per the criteria 
we defined. Using the Rest API provided by GitHub we collected information on the issues for 
all these NPM packages. We collected the total number of issues for all these packages 
from 2015-03-01 to 2018-08-31.

Finally, we used the issue creation dates to construct a dataset of the total number of 
issues per day. We used the total number of issues instead of the number of open issues 
because we are interested in the number of issues encountered by the users of the packages. 
Whether an issue is resolved or not depends on a number of factors, \emph{e.g.} the number of 
developers, the responsiveness of the developers, the number of packages managed by each developer, the complexity of the problem; most of which are unrelated to usage, so we decided using the total number of issues is a much more reasonable option.
This same issue data was used by our analysis of the 520 NPM packages, where we used the number
 of issues created for those packages during one month prior to the date the npm-miner data was 
collected.

\vspace{-20pt}
\subsubsection{NPM data preprocessing}

\begin{figure}[!t]
\centering
\includegraphics[width=0.9\linewidth]{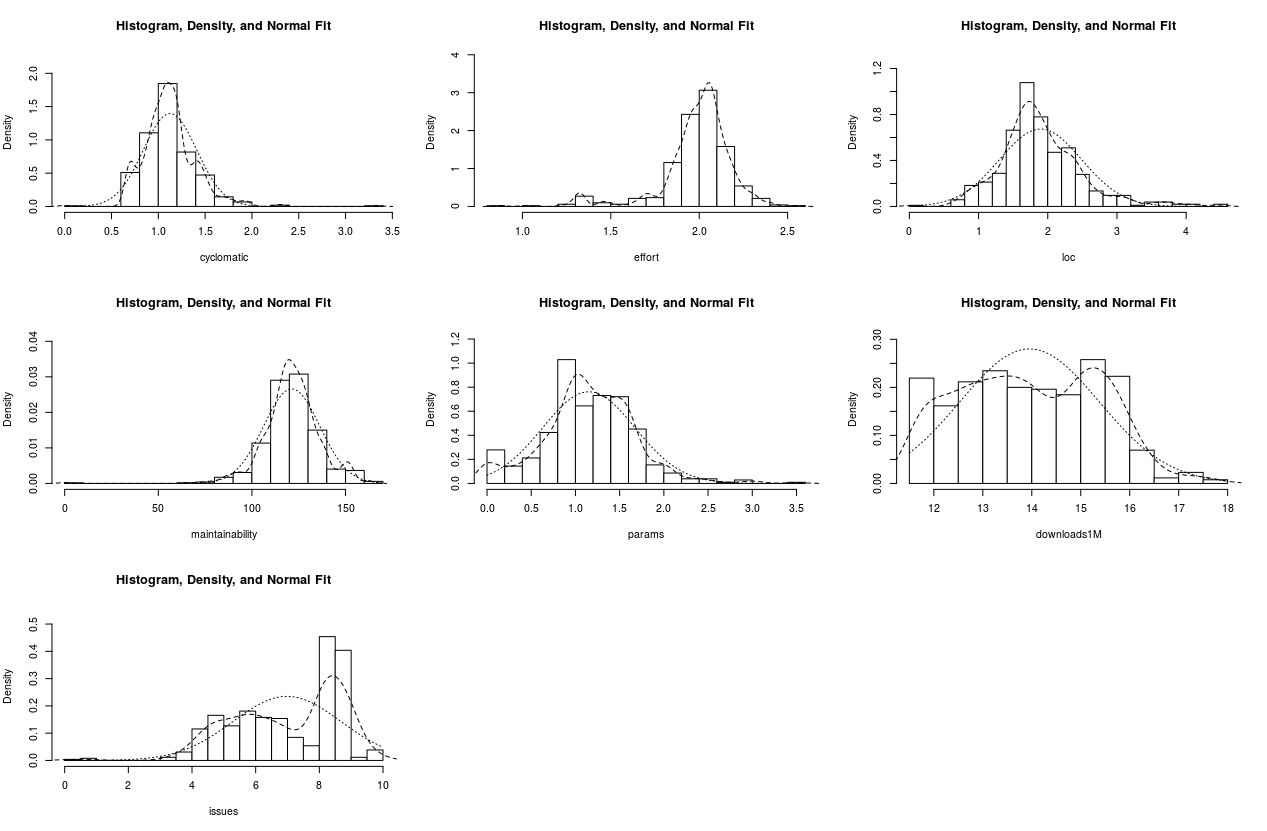}%
\caption{Distribution of the variables after transformation: the 520 NPM packages}
\label{fig:minerdist}
\vspace{-10pt}
\end{figure}

For the analysis of the 520 NPM packages, we constructed a dataset containing the 5 
quality measures of the packages (variable names: \textit{loc, cyclomatic, effort, 
params,} and \textit{maintainability} ), the number of downloads (variable name: \textit{downloads1M}) 
during the month before the data collection date for npm-miner dataset (2018-01-22), and the 
number of issues created for those packages during the same time. The variables were log 
transformed to correct the skewness of the data, except \textit{maintainability} and 
\textit{effort}, since these variables were not skewed. The distribution for the transformed 
data is shown in Figure~\ref{fig:minerdist}. For the purpose of applying BN models, 
the data was scaled as well. 

The data for the 4430 NPM packages which was used to compare the trends had the calendar date, 
the cumulative number of issues for the packages until that date, and the number of downloads 
on that date. Since this data was used for demonstrating the trends, we did not transform this 
dataset.

\section{Methodological Overview}\label{s:method}

In this section we describe the methodology we followed in this paper. We employed two 
different modeling techniques for finding out the relationship among the post-release 
variables: (1)  Bayesian structure search method, and (2) Random Forest Regression method. 
Since we primarily focus on finding which variables have the most impact on the number of 
exceptions (or issues for the NPM packages), we used it as the response variable for the Random 
Forest regression model. 

We considered using the OLS estimator since it is one of the simplest modeling methods that 
gives good models in a lot of situations and the result is easy to interpret. However, we were
 unsure about the accuracy of the result due to the presence to moderate to high correlation 
between some of the predictor variables (e.g. the Release Date and Release Duration variables 
had a correlation of -0.88 for the iOS application data). Moreover, we found that our variables 
do not satisfy all the criteria (laid out by the Gauss–Markov theorem) for creating the  best 
linear unbiased estimator (BLUE),  so we ended up not using it in our study.
Instead, we decided to use Bayesian Network (BN) for modeling the interrelationship among these
 variables, since the accuracy of this model is unaffected by the presence of high correlation 
among the predictors. Variables with high correlation simply appear as connected nodes in the
 final model. This eliminates the need of dropping some of the correlated variables from the 
model, which introduces subjectivity during the modeling process.
Since the use of Bayesian Network models is not very common in this context, we discuss BN 
models in greater detail later in this section. The other modeling approach we used is Random 
Forest regression method. Random Forest is one of the best off-the-shelf models that works well 
with almost all types of data and generally does not overfit,  and it is easy to get the 
relative importance of the predictor variables from a fitted model. These two factors led us to 
use Random Forest regression as the other modeling technique to identify the most impactful 
predictors explaining the number of exceptions. To find the best fitting Random Forest model, 
we performed a grid search using the ``tune'' function of the ``e1071'' R package to find the 
best model parameters ``ntree'': the number of tress to grow, and ``mtry'': the number of 
variables randomly sampled as candidates at each split. Since the sample size of datasets are 
limited, we used 10 times 2 fold cross-validation as the tuning method.

\vspace{-15pt}
\subsection{Bayesian Network Models}
Bayesian Network~\cite{koller2009probabilistic,scutari2010introduction} 
is a type of Probabilistic Graphical Model (PGM), which explicitly represents the
conditional dependency/independence as a directed acyclic graph where variables
represent nodes and dependencies represent links, and thus this
representation can be used as a generative model\footnote{A
  generative model specifies a joint probability distribution over
  all observed variables, whereas a discriminative model
  (like the ones obtained from regression or decision trees) provides a
  model only for the target variable(s) conditional on the predictor
  variables. Thus, while a discriminative model allows only sampling
  of the target variables conditional on the predictors, a
  generative model can be used, for example, to simulate
  (i.e. generate) values of any variable in the model, and
  consequently, to gain an understanding of the underlying mechanics
  of a system, generative models are essential.}.  
Bayesian Networks models can be useful in the context of Software Engineering research~\cite{fenton1999critique} due to having several advantages over
regression models. To be precise, regression analysis is a very
simple BN where there is one directed link from each independent variable
to dependent variable. BNs, therefore, can help with 
multicollinearity, a common problem with software engineering 
data~\cite{yu2002predicting,subramanyam2003empirical,briand2000exploring,Changes07}, 
that is present in our data as well, by linking independent variables. 

Another variety of PGM that we did not use in this paper 
(details in Section~\ref{s:limitation}) is the Markov random
fields that represent the interrelationships between variables 
as undirected graphs. They differ in the set of
independencies they can encode and the factorization of the
distribution that they induce~\cite{koller2009probabilistic}.

\noindent\\
\textbf{Bayesian Network Model construction:}
Despite the promises of BNs, they tend to be quite sensitive to data,
and operational data, is often problematic~\cite{M14,zmz15}. 
Careful preprocessing, therefore, is needed to ensure a 
reliable and reproducible result. Two primary ways to use BNs exist. With the first
approach the graph represents dependencies obtained from domain experts.
The graph may include prior distributions
about the parameters of the overall model. The data is then used to
calculate the posterior distribution and to make inference. The
second approach puts minimal a-priori assumptions about the model
and focuses on the search for the best graphical representation for a given dataset
(structure learning). This is an NP-hard problem~\cite{chickering1996learning},
but a number of different heuristic structure learning algorithms are
available. Due to the lack of any strong theory connecting the variables we are considering,
we decided to use the structure search method for BN model construction.
Since our goal is to find a Bayesian network model for the data, we didn't
examine the methods that do not result in a Directed Acyclic Graph
(DAG). We found that the \textit{bnlearn} package in R implements a wide range of BN searching
methods for continuous, discrete, or a mixed set of variables and
the corresponding families of scoring functions and also has a good number of examples.
These methods were also shown to be able to recover the underlying network for a 
protein-signaling-chain (in Biology) in~\cite{bnppt}. We, therefore,
use this package for our analysis. In addition to the methods
implemented in \textit{bnlearn} package, we investigated some 
methods from a few other packages which can be interfaced with
the \textit{bnlearn} package.

Due to the potential inconsistencies of the BN models, we performed our 
modeling in two stages. First, we considered all available BN structure 
methods in the \textit{bnlearn} package and ran a simulation based study to 
find the methods that are most accurate and then we used those methods on our 
data to create the final model.

\noindent\\
\textbf{Methods considered:}\\
The different BN structure search methods we considered are listed below:

\begin{itemize}
\item \textit{Greedy Hill-Climbing search algorithms}(HC)~\cite{nagarajan2013bayesian,bnppt}
\item \textit{Hybrid algorithms}(Hybrid)~\cite{nagarajan2013bayesian,bnppt}
\item \textit{Posterior maximization} using \textit{deal} package in R~\cite{bnppt,dealR} .
\item \textit{Simulated Annealing} using \textit{catnet} package in R~\cite{catnetR,bnppt}.
\item \textit{PC Algorithm} using \textit{pcalg} package in R~\cite{pcalgR,pcalgR2,bnppt}.
\item \textit{MAP (maximum a-posteriori estimation) Bayesian Model Averaging} (MAP) ~\cite{nagarajan2013bayesian,bnppt}
\end{itemize}

This is not an exhaustive list of all possible BN structure search methods, in fact,
it is impossible to make an exhaustive list for a heuristic search method like this, however, 
they represent a class of popular heuristic structure search methods that are part of the
``bnlearn'' package, which is a popular R package that is in continuous development since 2007.

All structure search algorithms try to maximize some form of a network score. 
Among the various scores available, BIC score is the suitable one when 
the goal is to create an explanatory model from non-informative prior models
~\cite{shmueli2010explain,sober2002instrumentalism}.
BIC score is used for discrete data while the Gaussian 
equivalent of BIC (bic-g) score is used for continuous data.

The results, \emph{i.e.} the structure and the parameters resulting from a structure
search algorithm, are often noisy, meaning that
different settings induce slightly different networks. To mitigate
this effect we use non-parametric bootstrap model averaging method
described in~\cite{friedman1999data}, which provides confidence
level for both the existence of edge and its direction. This enables
us to select a model based a confidence threshold. Authors
of~\cite{friedman1999data} argue that threshold is domain specific
and needs to be determined for each domain. For instance, a threshold of 
 0.95 indicates that  only the edges that appeared in more than
 95\% of the bootstrap optimized models were selected. 
 
Many applications of BNs discretize the data prior to applying
the structure learning methods, and in some cases where the data distribution is too skewed to fit the normality assumption, discretizing the data produces better models than using continuous data, so we considered it as a possibility as well.

Using continuous data works best
when the random variables (possibly after a transformation) 
have Gaussian distribution. While using
discrete data does not require such assumptions, obtaining the optimal
discretization for a dataset is in itself an NP-hard
problem~\cite{chlebus1998finding}. Choosing a
sub-optimal discretization technique may result in spurious or
missed relationships, which can in turn result in incorrect
dependencies being reported in the resulting model. Given the pros
and cons of both types of methods, we use methods of both types for
our simulation study. As we are interested in creating a generative model, we had to
use a discretization method that is unsupervised. The basic problem
with commonly used supervised methods (\textit{e.g.} Chi-square, or
MDLP discretization algorithms) is that they optimize
discretization to improve explanatory power for 
a single response variable. This is not suitable for a 
BN structure search, because we do
not know  which variables will be responses (have arrows
pointing to them) and which will be independent (have no incoming
arrows) {\em a-priori}. While some research on multidimensional
discretization methods exists~\cite{perez2006supervised},
we are not aware of any that have a robust implementation.

\noindent\\
\textbf{Simulation Study:}\\
\begin{figure}
\centering
\includegraphics[width=0.65\linewidth]{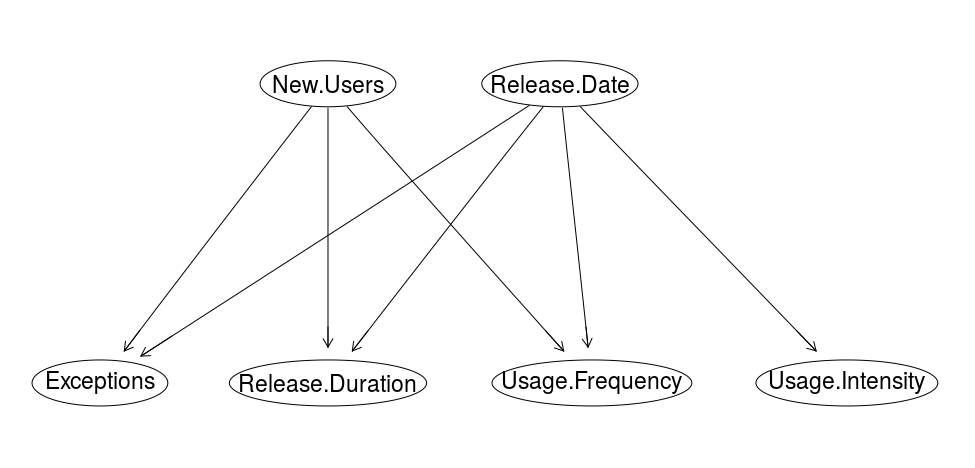}
\caption{Custom model used for Simulation Study}
\label{fig:theory}
\vspace{-15pt}
\end{figure}
We performed the simulation study by first creating 
a random BN (see Figure~\ref{fig:theory}) with 
six nodes, since we also have six variables in our final list (Table~\ref{t:finalvarss}).
For demonstration purposes  we use the same variable names. 
We fitted this graph with our data on GA releases of Avaya communicator for Android
 (log-transformed and scaled) to generate values for the coefficients 
for each edge. This model was used in our simulation study going forward.
We created 1000 different simulated datasets from the BN structure in Figure~\ref{fig:theory},
and applied the different structure search algorithms 
(both continuous and discrete versions, where available) listed above. Our performance metric
is finding how many times the different algorithms can recover the 
underlying structure from the simulated data. 

Other than testing the methods  themselves, we also tested whether or not we should 
discretize the data. We tried different discretization methods, \textit{viz.} 
equal interval, equal frequency, and
k-means clustering based discretization methods from the
\textit{arules} package~\cite{arulesR}, and the 
Hartemink\footnote{Hartemink's pairwise mutual information 
method\cite{hartemink2001principled}.} discretization methods 
in the \textit{bnlearn} package.

Except for the \textit{Posterior maximization} using \textit{deal} package, which can't be bootstrapped, 
all other results were bootstrapped, so we tested different thresholds in our 
simulation study as well. Finally, for the the Hybrid search algorithm, in which
conditional independence tests are performed to restrict
the search space for a subsequent greedy search, there are many restrict methods
available, \textit{viz.} gs" (Grow-Shrink), "iamb" (IAMB), "fast.iamb" (Fast-IAMB), "inter.iamb" (Inter-IAMB), "mmpc" (Max-Min Parent Children), "si.hiton.pc" (Semi- Interleaved HITON-PC), "chow.liu" (Chow-Liu), "aracne" (ARACNE)~\cite{bnlearnR}, and we tested all of these restrict options 
in our simulation study. 

\begin{table}
\parbox{.45\linewidth}{
\caption{Result of Simulation Study}\label{t:sim_result}
\resizebox{0.5\columnwidth}{!}{%
\begin{tabular}{lll}
\hline
\textbf{Method} & \textbf{Exact} & \textbf{Off-by-one} \\ \hline
HC & 0.574 & 0.264 \\
MAP & 0.596 & 0.214 \\
Hybrid- si.hiton.pc & 0.000 & 0.019 \\
Hybrid- mmpc & 0.000 & 0.016 \\
Hybrid- gs & 0.000 & 0.011 \\
HC-D-F & 0.000 & 0.010 \\
Hybrid- iamb & 0.000 & 0.010 \\
Hybrid- mmpc -D-H & 0.000 & 0.008 \\
Hybrid- si.hiton.pc -D-H & 0.000 & 0.008 \\
HC-D-H & 0.000 & 0.007 \\
Hybrid- mmpc -D-F & 0.000 & 0.007 \\
Hybrid- si.hiton.pc -D-F & 0.000 & 0.006 \\
Hybrid- iamb -D-F & 0.000 & 0.005 \\
Hybrid- gs -D-F & 0.000 & 0.004 \\
Hybrid- gs -D-H & 0.000 & 0.004 \\
Hybrid- iamb -D-H & 0.000 & 0.002\\\hline
\end{tabular}
}}
\hfill
\parbox{.45\linewidth}{
\caption{Result of Simulation Study: Different Thresholds}\label{t:sim_threshold}
\resizebox{.45\columnwidth}{!}{%
\begin{tabular}{llll}
\hline
Method & Threshold & Exact & Off-by-one \\ \hline
MAP & 0.85 & 0.68 & 0.25 \\
MAP & 0.80 & 0.67 & 0.25 \\
MAP & 0.90 & 0.67 & 0.26 \\
MAP & 0.95 & 0.66 & 0.27 \\
MAP & 1.00 & 0.66 & 0.27 \\
MAP & 0.75 & 0.66 & 0.21 \\
HC & 0.65 & 0.63 & 0.23 \\
HC & 0.70 & 0.63 & 0.23 \\
HC & 0.75 & 0.63 & 0.23 \\
HC & 0.80 & 0.63 & 0.23 \\
HC & 0.85 & 0.62 & 0.24 \\
HC & 0.55 & 0.62 & 0.23 \\
HC & 0.60 & 0.62 & 0.23 \\
MAP & 0.70 & 0.62 & 0.21 \\
HC & 0.90 & 0.60 & 0.26 \\
MAP & 0.65 & 0.58 & 0.17 \\
HC & 0.95 & 0.57 & 0.29 \\
MAP & 0.60 & 0.43 & 0.14 \\
MAP & 0.55 & 0.33 & 0.11 \\
HC & 1.00 & 0.19 & 0.47\\ \hline
\end{tabular}
}}
\vspace{-10pt}
\end{table}

The result of the simulation study is shown in Table~\ref{t:sim_result}, which shows 
the fraction of times exact structures and off-by-one structures\footnote{one extra 
/ missing / reversed edge} were generated by each method in the simulation. The result varies 
with the chosen threshold, so in Table~\ref{t:sim_result}, we show the overall performance of the
different methods  which generated an exact or off-by-one structure at least once in the simulation.
For the hybrid search methods, we list mention the restrict option that was used, and the 
`-D' suffix indicates a discretization method was used to discretize the data prior to applying 
a structure search method. `-D-H' indicates Hartemink discretization method and `-D-F' indicates 
Equal-Frequency discretization method. It is clear from the table that only HC and MAP methods
can effectively reproduce the correct underlying structure around half of the times and they create 
more off-by-one structures than others, indicating the error rate is the lowest for these methods.

In Table~\ref{t:sim_threshold}, we show the fraction of times exact and 
off-by-one models were generated by HC and MAP methods, which performed 
the best among the methods considered,  for different thresholds. 
It can be seen that using a moderately high threshold between 0.75 and 0.9 gives
good results for both HC and MAP, while higher thresholds for HC and lower thresholds for MAP
give worse results. Using the optimal threshold creates models that have more than one wrong 
and/or missing edge only 7-14\% of the times. 

The result of the simulation study had the following findings:
\begin{itemize}
\item Using structure search algorithms on the continuous data resulted in much more frequent recovery of the original BN structure compared to discretized data.
\item Bootstrapping  improves the stability of the results considerably.
\item The bootstrapped Hill-Climbing search and MAP Bayesian Model Averaging algorithms outperformed all others both in terms of accuracy and runtime, being able to recover the underlying structure more than 63\% of the times and making no more than one error 86\% of times with optimal thresholds. 
\end{itemize}

\emph{We consider this study one of the contributions of the paper, and hope that it 
would be useful for researchers using BN structure learning techniques.}

\vspace{-10pt}
\section{Answering the Research Questions: Results and Analysis}\label{s:result}

\vspace{-10pt}
\subsection{RQ1: Modeling the relationship between Exceptions and other post-release variables}\label{s:explain}
As mentioned earlier, we conducted our analysis in two stages: first,  we used Bayesian Network (BN) modeling approach to identify the interrelationship between the variables and then, we used a random forest (RF) model to verify the results.

\vspace{-10pt}
\subsubsection{Bayesian Network Model}
One key assumption for applying the continuous BN structure search algorithms 
is that the variables have a distribution close to a Gaussian distribution.
To satisfy this modeling assumption, we scaled all the
variables to unit scale. The variable ``Exceptions" still had a long
tailed distribution, but the distributions of the other variables
were much closer to normal distribution.

\begin{figure}[!t]
\centering
\includegraphics[width=0.7\linewidth]{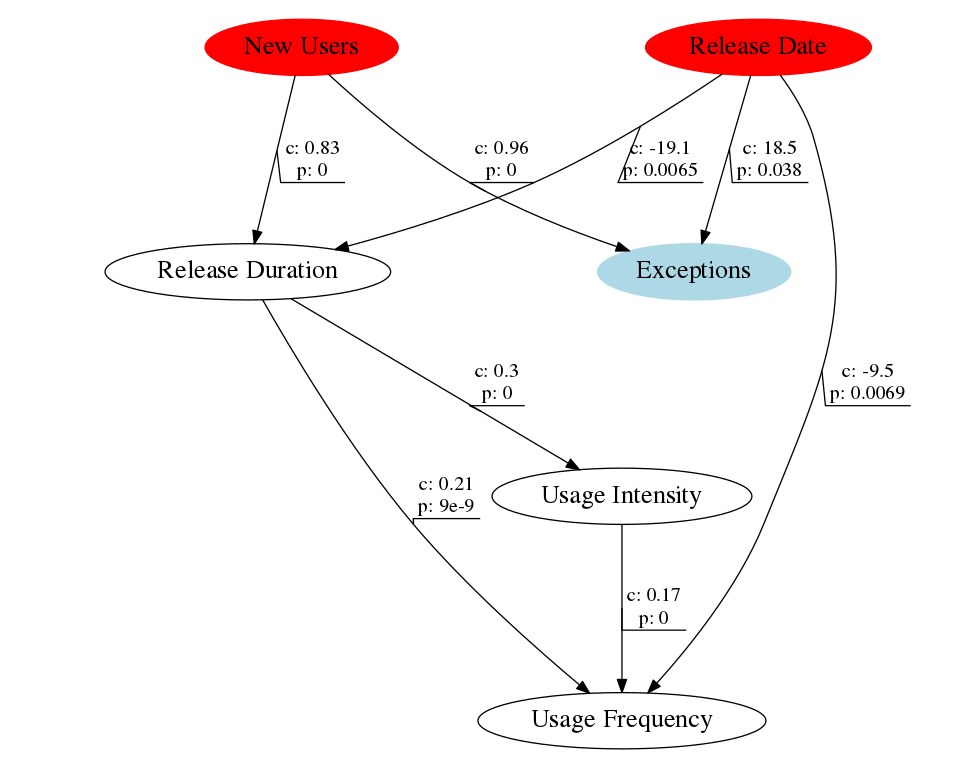}%
\caption{Final BN Model for GA releases of Avaya Communicator for Android (with c: coefficients after fitting the transformed, but unscaled data, p: p-value  for the link) }
\label{fig:finalAGA}
\vspace{-10pt}
\end{figure}

\begin{figure}[!t]
\centering
\includegraphics[width=0.65\linewidth]{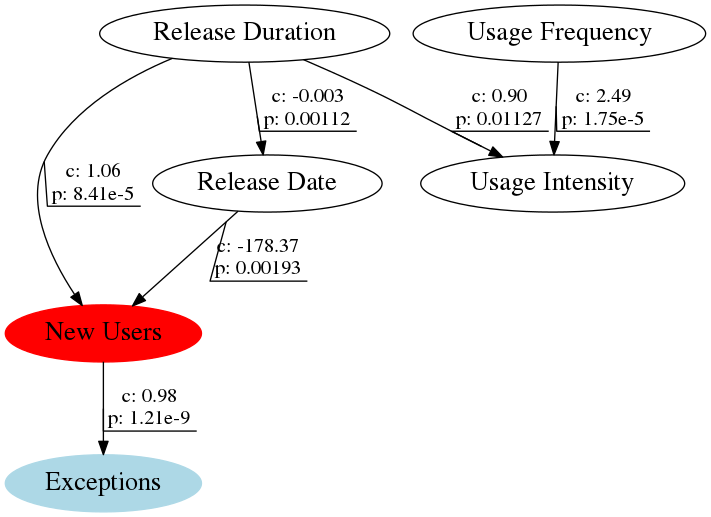}%
\caption{Final BN Model for Development releases of Avaya Communicator for Android (with c: coefficients after fitting the transformed, but unscaled data, p: p-value  for the link) }
\label{fig:finalAD}
\vspace{-10pt}
\end{figure}

\begin{figure}[!t]
\centering
\includegraphics[width=0.7\linewidth]{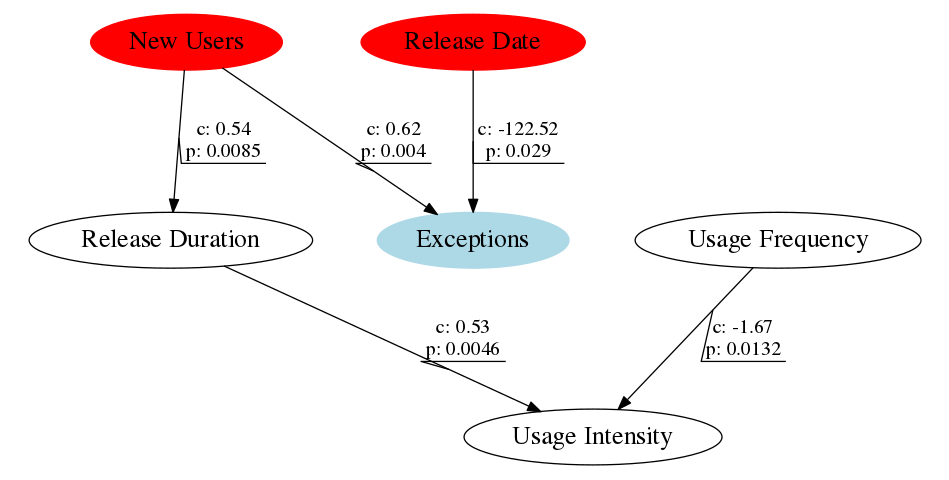}%
\caption{Final BN Model for GA releases of Avaya mobile SIP for iOS (with c: coefficients after fitting the transformed, but unscaled data, p: p-value  for the link) }
\label{fig:finalI}
\vspace{-10pt}
\end{figure}

According to the result of the simulation study, we decided to use bootstrapped hill-climbing search and MAP Bayesian model averaging methods for constructing the Final BN models for our datasets and considered the model that resulted from both the methods. 
The resultant BN model for the GA releases of Avaya Communicator for Android is shown in Figure~\ref{fig:finalAGA}, which shows ``New.Users'' and ``Release.Date'' are parent nodes of ``Exceptions''.  Figure~\ref{fig:finalAD} shows the final BN Model for Development releases of Avaya Communicator for Android, in which only ``New.Users'' is the parent of ``Exceptions'', and Figure~\ref{fig:finalI} shows the final BN Model for GA releases of Avaya mobile SIP for iOS, where once again ``New.Users'' and ``Release.Date'' are parent nodes of ``Exceptions''. In these figures p-values $< 2e-16$ are denoted as 0. 

Every bootstrap run was performed over 500 bootstrap samples, and a
hill-climbing search  with 100 random restarts was applied on each sample 
to find the best fitting network, so in essence, each resultant network was 
obtained by averaging 50,000 candidate networks. We used a Threshold of 
0.85, as it seemed optimal from our simulation study.

\begin{table}[ht]
\caption{Example bootstrap result - GA releases of Avaya Communicator for Android}\label{t:boot}
\centering
\resizebox{0.65\columnwidth}{!}{%
\begin{tabular}{llrr}
  \hline
 from & to & strength & direction \\ 
  \hline
Exceptions & New.Users & 1.00 & 0.34 \\ 
  Exceptions & Release.Date & 0.86 & 0.47 \\ 
  Exceptions & Release.Duration & 0.46 & 0.50 \\ 
  Exceptions & Usage.Frequency & 0.75 & 0.78 \\ 
  Exceptions & Usage.Intensity & 0.35 & 0.47 \\ 
  New.Users & Exceptions & 1.00 & 0.66 \\ 
  New.Users & Release.Date & 0.20 & 0.62 \\ 
  New.Users & Release.Duration & 1.00 & 0.71 \\ 
  New.Users & Usage.Frequency & 0.71 & 0.85 \\ 
  New.Users & Usage.Intensity & 0.34 & 0.64 \\ 
  Release.Date & Exceptions & 0.86 & 0.53 \\ 
  Release.Date & New.Users & 0.20 & 0.38 \\ 
  Release.Date & Release.Duration & 1.00 & 0.63 \\ 
  Release.Date & Usage.Frequency & 0.97 & 0.82 \\ 
  Release.Date & Usage.Intensity & 0.66 & 0.77 \\ 
  Release.Duration & Exceptions & 0.46 & 0.50 \\ 
  Release.Duration & New.Users & 1.00 & 0.29 \\ 
  Release.Duration & Release.Date & 1.00 & 0.37 \\ 
  Release.Duration & Usage.Frequency & 0.90 & 0.55 \\ 
  Release.Duration & Usage.Intensity & 1.00 & 0.53 \\ 
  Usage.Frequency & Exceptions & 0.75 & 0.22 \\ 
  Usage.Frequency & New.Users & 0.71 & 0.15 \\ 
  Usage.Frequency & Release.Date & 0.97 & 0.18 \\ 
  Usage.Frequency & Release.Duration & 0.90 & 0.45 \\ 
  Usage.Frequency & Usage.Intensity & 1.00 & 0.22 \\ 
  Usage.Intensity & Exceptions & 0.35 & 0.53 \\ 
  Usage.Intensity & New.Users & 0.34 & 0.36 \\ 
  Usage.Intensity & Release.Date & 0.66 & 0.23 \\ 
  Usage.Intensity & Release.Duration & 1.00 & 0.47 \\ 
  Usage.Intensity & Usage.Frequency & 1.00 & 0.78 \\ 
   \hline
\end{tabular}
}
\vspace{-10pt}
\end{table}

The result form a bootstrap run shows the relative strength of the
link and the relative confidence for the direction of the link. 
In Table~\ref{t:boot} we have shown the result from one bootstrap run of the HC method for all possible edges for the GA release data of Avaya Communicator for Android. If an edge has $<50\%$ confidence in its direction, then the edge appears in the opposite direction in our model.
Although Bayesian Networks are sometimes interpreted as causal relationships~\cite{pearl2011bayesian}, there are disagreements on how that should be done.
We, therefore, are not interpreting these relationships as causal here. All observed links, therefore, indicate the presence of observed correlation (and are empirical in nature) and the direction is a property of the topological ordering of nodes in a DAG, and affects the total probability distribution of the variables.

The BN models were fitted to
the unscaled data, and the resulting coefficient of each link is also shown
in the figures. The p-value for each link was calculated from a
linear model with the source nodes as predictors and the destination
node as the response variable, e.g. the p-value for the link from
``New.Users" to ``Exceptions" was calculated by looking at the
result of:  \texttt{  lm(Exceptions $\sim$ New.Users $+$ Release.Date)}. \\
We fitted the model to the transformed, but unscaled data (for easier interpretation of results). 

By looking at the p-values for the links, we can say that all the links in the BN models
are statistically significant. 
Links having a negative coefficient indicate an inverse relationship between the parent 
and the child node. The performance of explanatory models is evaluated by the fraction
of deviance explained by the model. Our model explains $80.3\%$ and $45.9\%$ of the
variation in ``Exceptions" (adjusted $R^2$ value of the model) for development and GA releases for Avaya Communicator for Android respectively and $42.0\%$ for GA releases of Avaya mobile SIP for iOS. This indicates our BN model is statistically significant, but the predictors we used could only explain around half of the variance in Exceptions.

\vspace{-10pt}
\subsubsection{Random Forest Model}
\begin{figure}[!t]
\centering
\begin{minipage}{.45\textwidth}
\includegraphics[width=\linewidth]{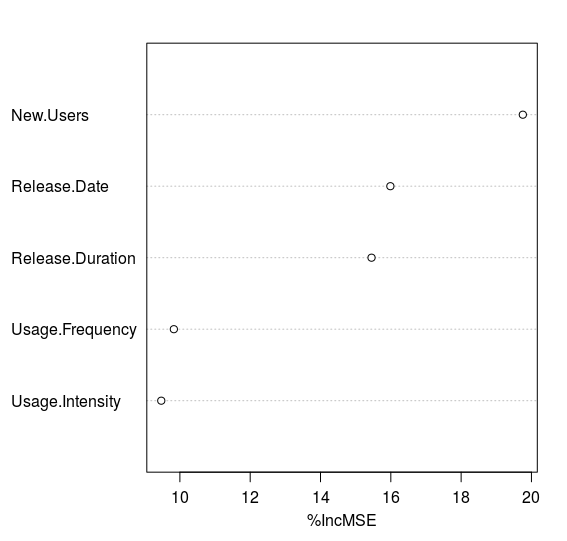}%
\caption{Variable Importance Plot of RF model for ``Exceptions" for GA release data of Avaya Communicator for Android}
\label{fig:rfAGA}
\end{minipage}
\hfill
\begin{minipage}{.45\textwidth}
\includegraphics[width=\linewidth]{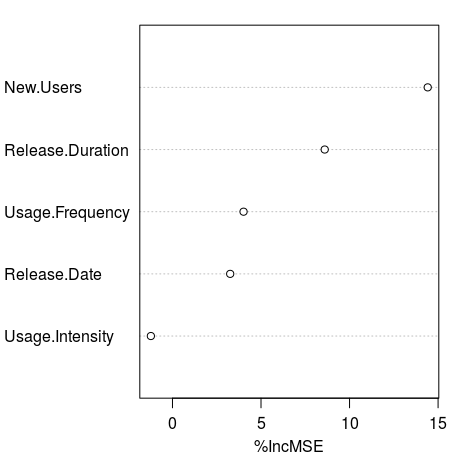}%
\caption{Variable Importance Plot of RF model for ``Exceptions" for development release data of Avaya Communicator for Android}
\label{fig:rfAD}
\end{minipage}
\vspace{-10pt}
\end{figure}

\begin{figure}[!t]
\centering
\includegraphics[width=0.4\linewidth]{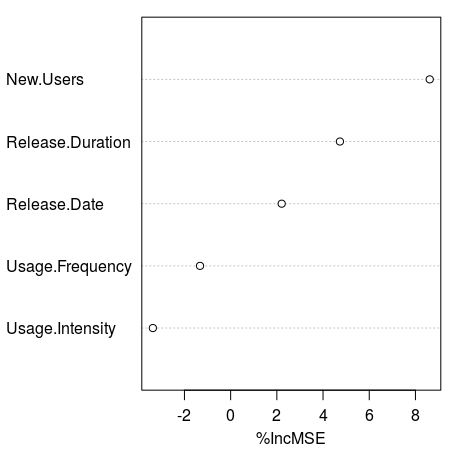}%
\caption{Variable Importance Plot of RF model for ``Exceptions" for GA releases of Avaya mobile SIP for iOS}
\label{fig:rfI}
\vspace{-10pt}
\end{figure}

As a verification step to identify the important variables affecting the number of exceptions, we used a Random Forest model to fit the data, with ``Exceptions'' as the response variable. The variable importance plot for the GA release data of Avaya Communicator for Android, as shown in Figure~\ref{fig:rfAGA}, indicates that ``Release.Date" and ``New.Users'' are the two most important variables. For the development releases of Avaya Communicator for Android, the variable importance plot is shown in Figure~\ref{fig:rfAD}. ``New.Users'' is again the most important variable, followed by ``Release.Duration''. For the GA releases of Avaya mobile SIP for iOS, the variable importance plot again shows the number of 
new users is the most important variable, as can be seen from Figure~\ref{fig:rfI}.

The best selected model parameters derived from tuning  show that the optimal models were obtained for ``ntree''$=$600 and ``mtry''$=$3 for all datasets. The $R^2$ values for these models, again obtained from 10 times 2 fold cross-validation, are $0.48$, $0.56$, and $0.31$ for the GA and development releases of the Android application and the GA releases of the iOS application respectively. The poor performance of the iOS application is likely due to the very small sample size of the dataset. Although the overall performance of the models wasn't very good, since we had a limited number of predictors, and none of the internal factors were part of the model, this result shows that even for the purpose of prediction, the number of new users play an important role.

\vspace{-15pt}
\subsection{RQ2: Deriving a usage-independent measure of Quality}\label{s:qual}

\vspace{-10pt}
\subsubsection{Obtaining the Quality measure}
In order to arrive at the usage independent quality measure, we follow the framework of establishing laws governing relationships among measures of software development proposed in~\cite{mockuskeynote}. Law is an equivalent of invariance, i.e. a function of measures that is constant under certain conditions. In this case we want it to be constant for releases that have the same quality. First, the law requires a plausible mechanism and second, an empirical validation. Each new user may have a different type of phone, operating system, service provider, geographic region, and usage pattern. It is reasonable to assume that some of these configurations lead to software malfunction manifested as an exception. This provides us with a plausible mechanism on how precisely more new users of one release might generate more exceptions even if we have two releases of identical quality. 
We rely on our models (all of which show the number of software 
exceptions to be dependent on the number of users and on the software release date) to obtain empirical validation of this postulated mechanistic relationship. Therefore,
we arrive at the following software law that is applicable for the investigated context: the average number of 
exceptions experienced by each user should, therefore, be independent of usage and depend only on the qualities of a software release.

In this section we test the above evidence-based hypothesis and provide the result of an analysis with the \textbf{number of exceptions per user} as a response variable (``Quality'') representing software quality. 
\emph{This is actually a measure for faultiness, so a lower value of ``Quality" indicates the actual quality of the software perceived by end users is better.}

The value of the ``Quality'' variable (not log transformed) was seen to be varying between 0 and 10.85 (mean: 0.45, median: 0, standard deviation: 1.48) for the GA release data of Avaya Communicator for Android, between 0 and 22.83 (mean: 1.12, median: 0, standard deviation: 4.55) for development versions of the same, and for the GA releases of Avaya mobile SIP for iOS it varied between 0 and 0.5 (mean: 0.0488, standard deviation: 0.15) . 

\vspace{-10pt}
\subsubsection{Establishing the independence of the Quality measure and other usage related variables}

Similar to the previous analysis, we applied Bayesian Network search and 
Random Forest modeling approaches on the dataset containing this quality measure and the remaining variables, all of which were log-transformed. 

\begin{figure}[!t]
\centering
\includegraphics[width=0.6\linewidth]{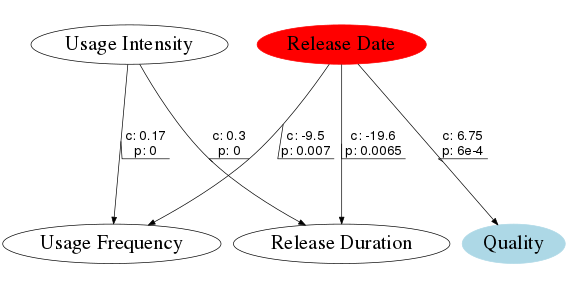}%
\caption{Bayesian Network  Model for ``Quality" - GA releases of Avaya Communicator for Android (with c: coefficients after fitting the transformed, but unscaled data, p: p-value  for the link)}
\label{fig:bn2AGA}
\vspace{-10pt}
\end{figure}

\begin{figure}[!t]
\begin{minipage}{.45\textwidth}
\raggedleft
\includegraphics[width=\linewidth]{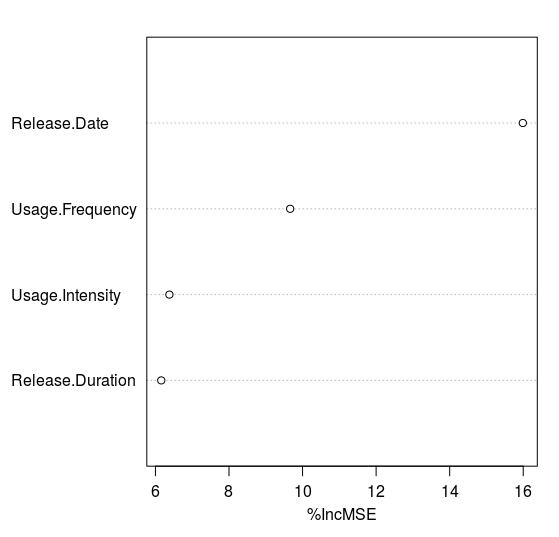}
\caption{Variable Importance plot from the Random Forest  Model for Quality Variable - GA releases of Avaya Communicator for Android}
\label{fig:rf2AGA}
\end{minipage}
\hfill
\begin{minipage}{.45\textwidth}
\raggedright 
\includegraphics[width=\linewidth]{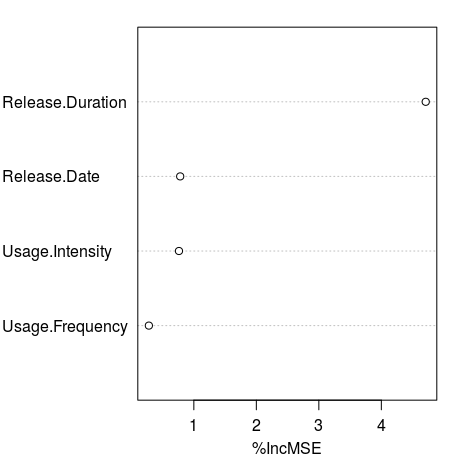}
\caption{Variable Importance plot from the Random Forest  Model for Quality Variable - Development releases of Avaya Communicator for Android}
\label{fig:rf2AD}
\end{minipage}
\vspace{-10pt}
\end{figure}

The result, as expected, shows that the quality of a software, measured by average number of faults experienced by each user, has no dependence on other usage variables. 
The BN model(Figure~\ref{fig:bn2AGA}), obtained with a threshold of 0.85 from a bootstrapped Hill-Climbing structure search model, indicates the ``Quality'' variable depends only on  the ``Release.Date'' variable.
Finally, the result of 10 times 2-fold cross-validation with the best RF model (Variable Importance plot in Figure~\ref{fig:rf2AGA}) with the optimal values for ``ntree''(300 in this case)) and ``mtry''(1 in this case)  indicates that the ``Release.Date'' variable is much more important compared to others, and the two usage related variables are of much lower importance. 

\begin{figure}[!t]
\centering
\includegraphics[width=0.6\linewidth]{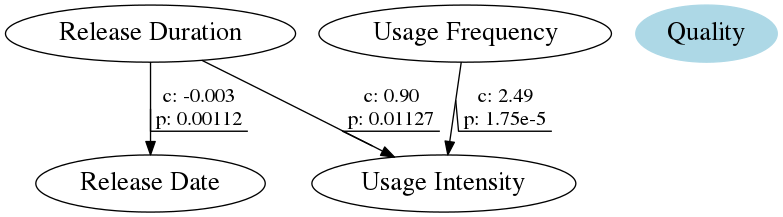}%
\caption{Bayesian Network  Model for ``Quality" - Development releases of Avaya Communicator for Android (with c: coefficients after fitting the transformed, but unscaled data, p: p-value  for the link)}
\label{fig:bn2AD}
\vspace{-10pt}
\end{figure}

For the development versions of Avaya Communicator for Android, all the predictor variables turned out be insignificant for the BN (Figure~\ref{fig:bn2AD}) models. Even the tuned RF model gives a really bad fit in the 10 times 2-fold cross-validation as well, with a $R^2$ value of -0.42 (the implication of a negative value of $R^2$ is as explained in~\cite{negRsq}), indicating the predictors are very poor. Still, the two usage related variables have the lowest importance in the variable importance plot as seen in Figure~\ref{fig:rf2AD}. 

\begin{figure}[!t]
\centering
\includegraphics[width=0.6\linewidth]{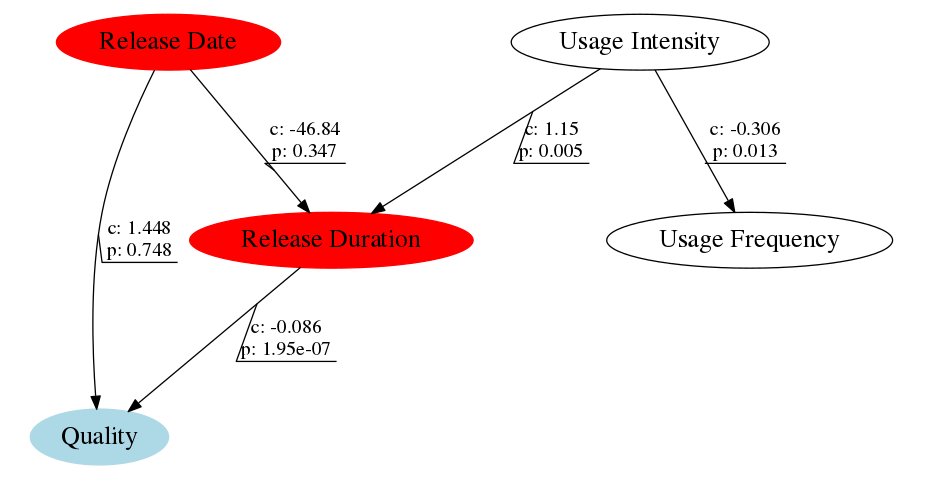}%
\caption{Bayesian Network  Model for ``Quality" - GA releases of Avaya mobile SIP for iOS (with c: coefficients after fitting the transformed, but unscaled data, p: p-value  for the link)}
\label{fig:bn2I}
\vspace{-10pt}
\end{figure}

Finally, for the GA releases of Avaya mobile SIP for iOS,  the BN model (Figure~\ref{fig:bn2I})  shows that release date and release duration have effect on the ``Quality'' variable, but the two other usage variables have no effect. We did not run RF model on this dataset owing to the very small sample size.

The results from these analyses clearly indicate that the quality measure defined by the number of exceptions per user is independent of software usage, and, therefore, suitable for comparing the quality of software development process among different releases of a software.

\vspace{-10pt}
\subsubsection{Timeline of Quality for the mobile Applications - RQ4}
\begin{figure}[!t]
\centering
\includegraphics[width=0.7\linewidth]{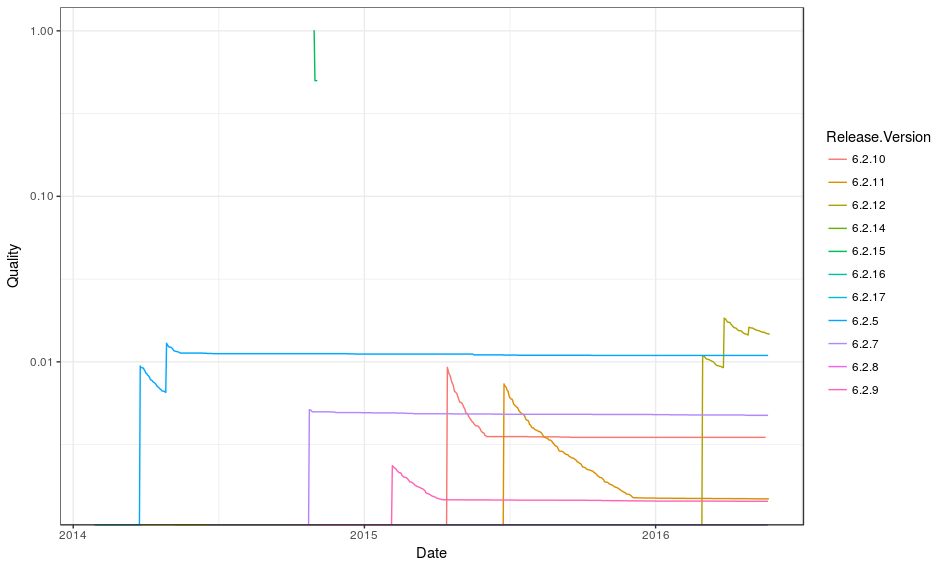}
\caption{Timeline for Quality Variable - GA releases of Avaya mobile SIP for iOS}
\label{fig:tI}
\vspace{-10pt}
\end{figure}

We wanted to see how the perceived quality of the releases of the different mobile applications described above change with time. As a general trend, we observe that most of the exceptions occur right after the release date. then, as the number of users keep increasing with time, the value of the quality variable drop and come to a stable value.  In this paper we show only the timeline for GA releases of Avaya mobile SIP for iOS (Figure~\ref{fig:tI}), since the other two softwares had a lot of releases, making them difficult to identify from the plot. The other two are  are available in our GitHub repository:\\ \url{https://github.com/tapjdey/release\_qual\_model}.

\begin{figure}[!t]
\centering
\includegraphics[width=0.8\linewidth]{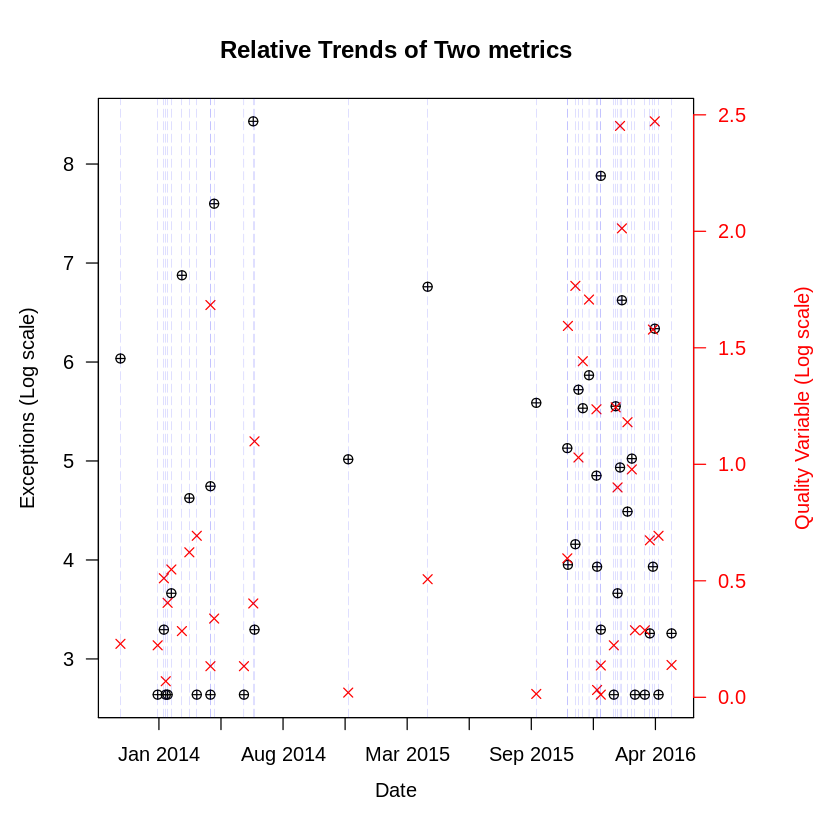}
\caption{Relative trends of Exception (marked with circle) and the Quality variable (marked with cross) - GA releases of Avaya Communicator for Android}
\label{fig:trend_ga}
\vspace{-10pt}
\end{figure}

\begin{figure}[!t]
\begin{minipage}{.45\textwidth}
\raggedleft
\includegraphics[width=\linewidth]{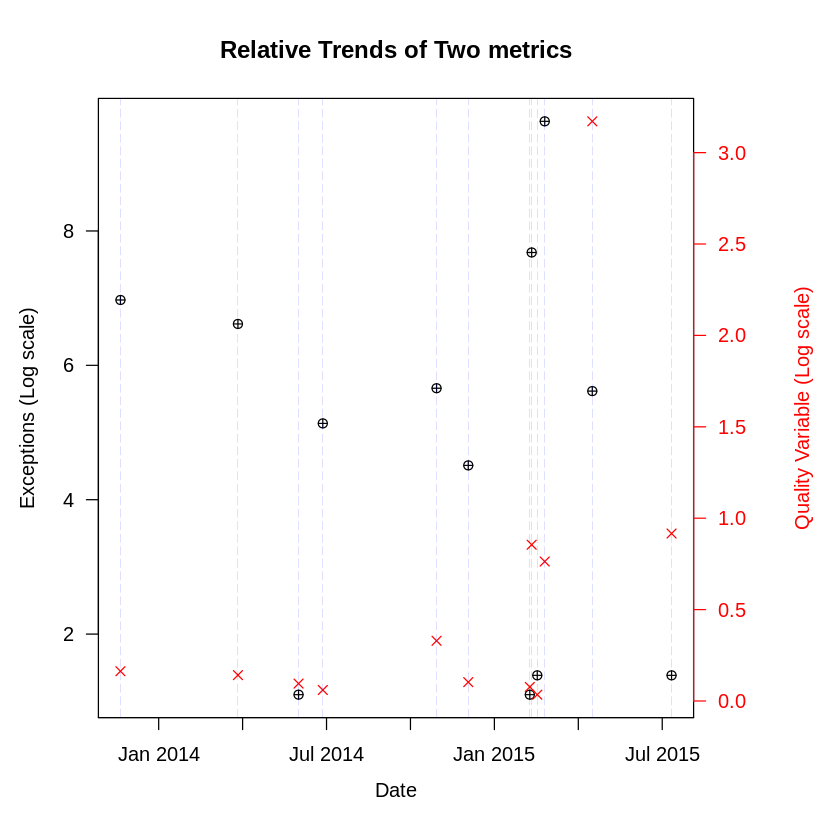}
\caption{Relative trends of Exception (marked with circle) and the Quality variable (marked with cross) - Development releases of Avaya Communicator for Android}
\label{fig:trend_dev}
\end{minipage}
\hfill
\begin{minipage}{.45\textwidth}
\raggedright 
\includegraphics[width=\linewidth]{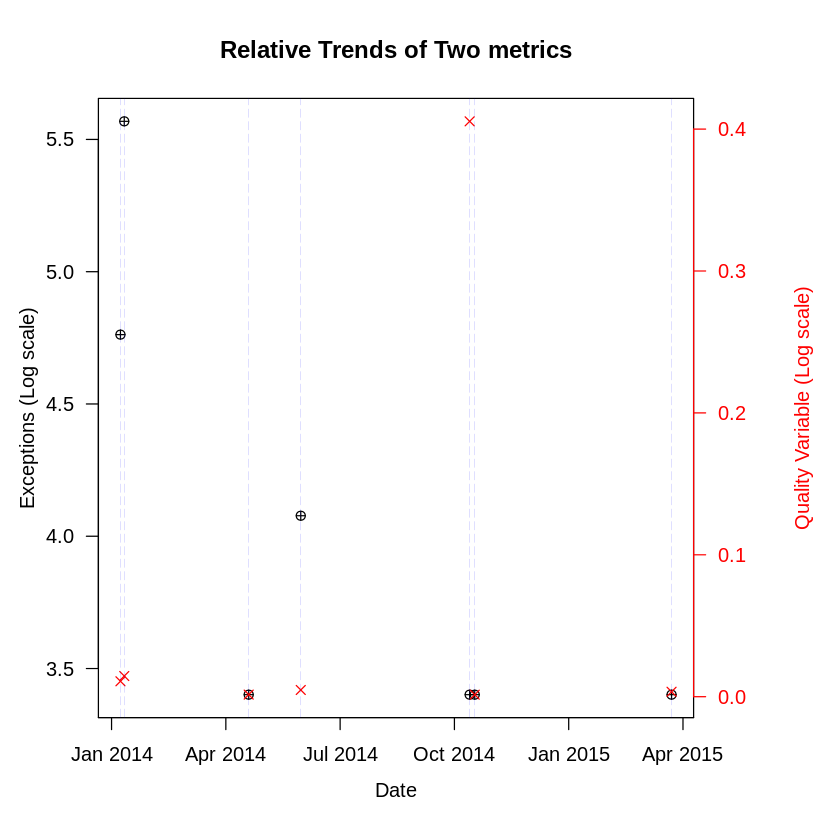}
\caption{Relative trends of Exception (marked with circle) and the Quality variable (marked with cross) - GA releases of Avaya mobile SIP for iOS}
\label{fig:trend_ios}
\end{minipage}
\vspace{-10pt}
\end{figure}

In Figure~\ref{fig:trend_ga},~\ref{fig:trend_dev}, and ~\ref{fig:trend_ios}, we show the relative trends of Exception and the Quality variable for the GA and development releases of the Android application and the GA releases of the iOS application respectively. We are not interested in the absolute values of the metrics, but the values of the metrics for a release relative to the values for other releases. We only show the releases with non-zero number of exceptions, since if the number of exceptions is zero, the the value of the quality metric is also zero. The blue dotted lines represent the releases dates of different releases, and the black marker and the red cross on the blue line represent the exceptions and the quality variable for that release respectively.

We can see that for a number of releases, Exceptions and Quality follow a similar trend, i.e. if the number of exceptions increase, the value of the Quality variable increases accordingly. However, there are indeed a number of cases where the number of exceptions is relatively small, but the value of Quality variable is larger than that for other releases, e.g. for many of the GA releases for the Android application in 2016, or vice versa, e.g. the release around September 2015 for the  GA releases for the Android application. This indicates that if we simply keep on using the number of exceptions as the quality measure, we will misclassify (as being better or worse than other releases) a number of releases. This result supports our hypothesis that not accounting for the usage parameter will not systematically misclassify all releases, since the internal factors affecting the release are independent of the external factors, but would randomly misclassify some of them depending on how much usage a release is getting.

\vspace{-10pt}
\subsection{RQ3: Analysis of the NPM Data and Results}\label{s:miner}

\begin{figure}[!t]
\begin{minipage}{.45\textwidth}
\raggedleft
\includegraphics[width=\linewidth]{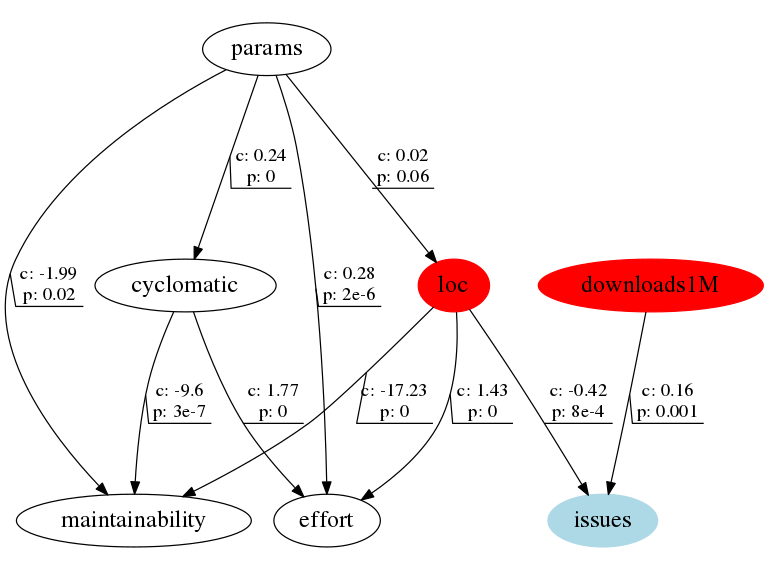}%
\caption{Bayesian Network  Model for Issues for the 520 NPM packages (with c: coefficients after fitting the transformed, but unscaled data, p: p-value  for the link)}
\label{fig:miner}
\end{minipage}
\hfill
\begin{minipage}{.45\textwidth}
\raggedright 
\includegraphics[width=\linewidth]{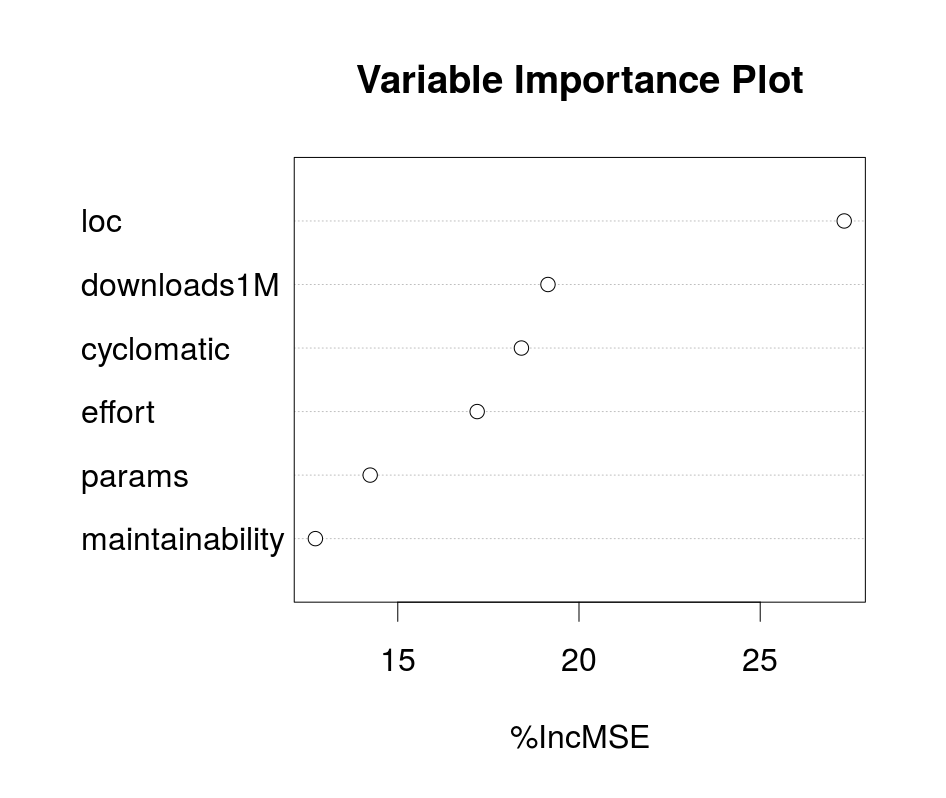}
\caption{Variable Importance plot for the Random Forest  Model for Issues for the 520 NPM Packages}
\label{fig:minervimp}
\end{minipage}
\vspace{-10pt}
\end{figure}

\begin{figure}[!t]
\begin{minipage}{.45\textwidth}
\raggedleft
\includegraphics[width=\linewidth]{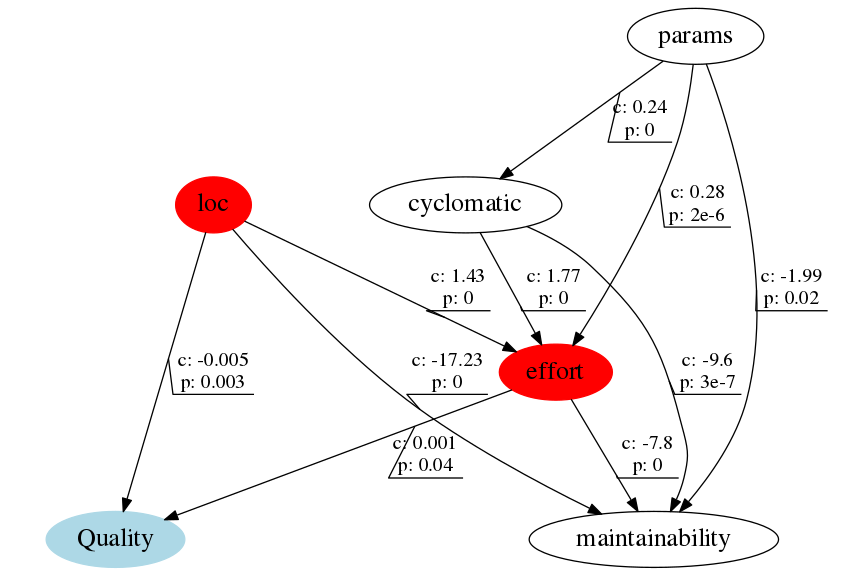}%
\caption{Bayesian Network  Model for Quality for the 520 NPM packages (with c: coefficients after fitting the transformed, but unscaled data, p: p-value  for the link)}
\label{fig:miner2}
\end{minipage}
\hfill
\begin{minipage}{.45\textwidth}
\raggedright 
\includegraphics[width=\linewidth]{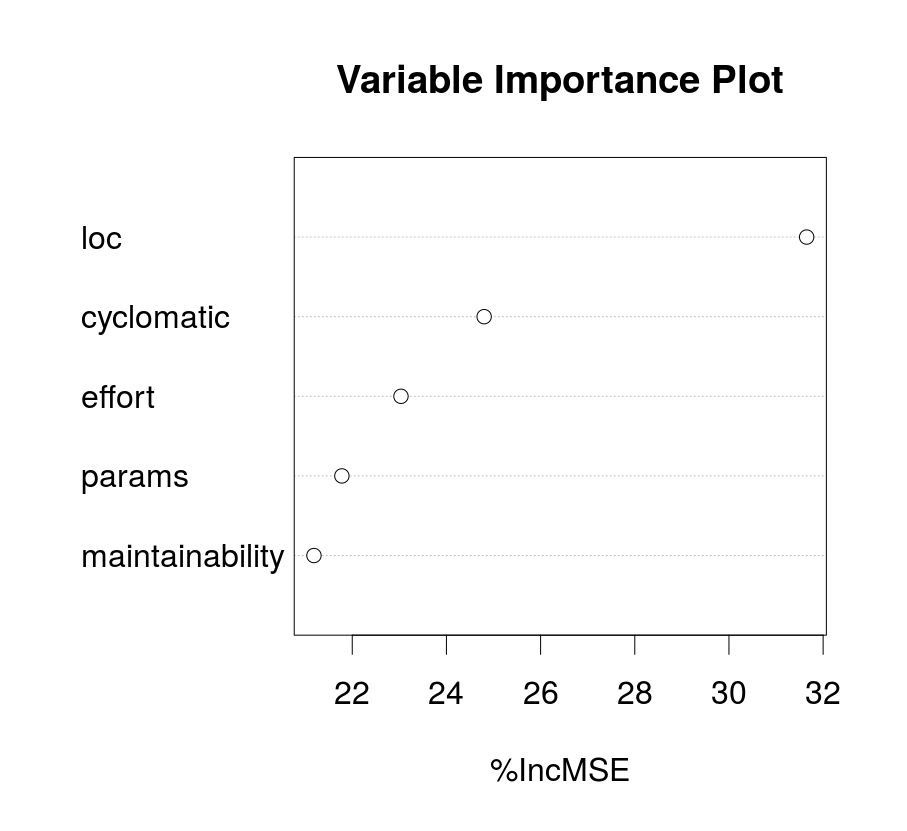}
\caption{Variable Importance plot for the Random Forest  Model for Quality for the 520 NPM Packages}
\label{fig:minervimp2}
\end{minipage}
\vspace{-10pt}
\end{figure}

We used the same modeling techniques as used in analyzing the mobile applications to 
examine the interrelationship between the various code complexity measures, extent of 
usage, and the number of issues. The BN model (Figure~\ref{fig:miner}) highlighted that the 
number of issues is dependent on the number of downloads as well as the number of logical 
lines of code. The value of $R^2$ for the BN model was found to be 0.59. 

The variable importance plot of the Random Forest model (Figure~\ref{fig:minervimp}) showed 
the number of logical lines of code to be the most important predictor, followed by the number 
of downloads over the month before data collection. The mean value of $R^2$, from performing a 
10 times 2 fold cross-validation was 0.64 (sd: 0.04) for this model. We wanted to see how the 
fit of the model changes if we drop the number of downloads from the list of predictors. This 
resulted in a significant drop in the value of $R^2$, which became 0.49 (sd: 0.02) under this 
condition. This result clearly indicates the importance of the number of downloads in modeling 
the number of issues even after taking the code complexity metrics into consideration.

Additionally, we decided to implement a usage-adjusted quality measure 
similar to what we had for the mobile 
applications and observe how it depends on the code complexity measures. 
Our quality measure (``Quality") was defined as the number of issues per download, similar to how
we defined it while analyzing the mobile applications.
The BN model (Figure~\ref{fig:miner2}) was similar to what we found last time, however, 
we now observed a new link from ``effort'' to ``Quality''. The fit of the model was found to 
be somewhat worse, with a $R^2$ value of 0.41. The Random Forest model gave a $R^2$ value of 
0.53 (sd:0.03) from a 10 times 2 fold cross-validation, and the variable importance plot 
(Figure~\ref{fig:minervimp2}) showed the lines of code to be the most impactful predictor.

\subsection{Timeline plots for NPM packages - RQ4}\label{s:npm}

We also presented the timelines for a few well-known NPM packages, showing the comparative trends of the number of issues and our proposed quality measure, defined as the number of issues per download. Since the number of downloads has a large variation, the quality measure also has a high degree of variation. So, we decided to fit a model to the quality variable, and add another line representing the fitted values of the model. We first tried to use a OLS model, but given the apparent non-linearity of the data, later decided to use a Generalized Additive Model (GAM). We did not fine tune this model, because it was only used for demonstrating the trend of the quality measure in the timeline plots.

Since the detailed analysis was performed on single releases of 520 packages, we wanted to 
verify if the number of downloads is an important predictor for the number of issues for all 
4430 packages individually during their lifetime for all releases. We found that out of the 
4430 packages, only for 36 packages the p-value of the predictor variable was more than 0.05 
before adjusting for the calendar date, and after adjusting for the calendar date, p-value was 
less than 0.5 (\emph{i.e.} the predictor was deemed significant) for all 4430 packages. 

The $R^2$ values of the fitted models varied between 0 and 0.8637 (median: 0.4982, standard 
deviation: 0.0618) before adjusting for the calendar date , and between 0.019 and 0.999 
(median: 0.9195, standard deviation: 0.0618) after  adjusting for the calendar date. So, 
we can say that the number of daily downloads is a significant and important predictor for 
the number of issues encountered by the users for most of the packages and the effect is more 
pronounced when the effect of automated downloads is controlled by calendar date. 

Since we just established that the number of issues of an NPM package depends on the number of daily downloads, a similar quality metric of the number of issues per download should be applicable in this situation as well. 

\begin{figure}[!t]
\centering
\includegraphics[width=0.75\linewidth]{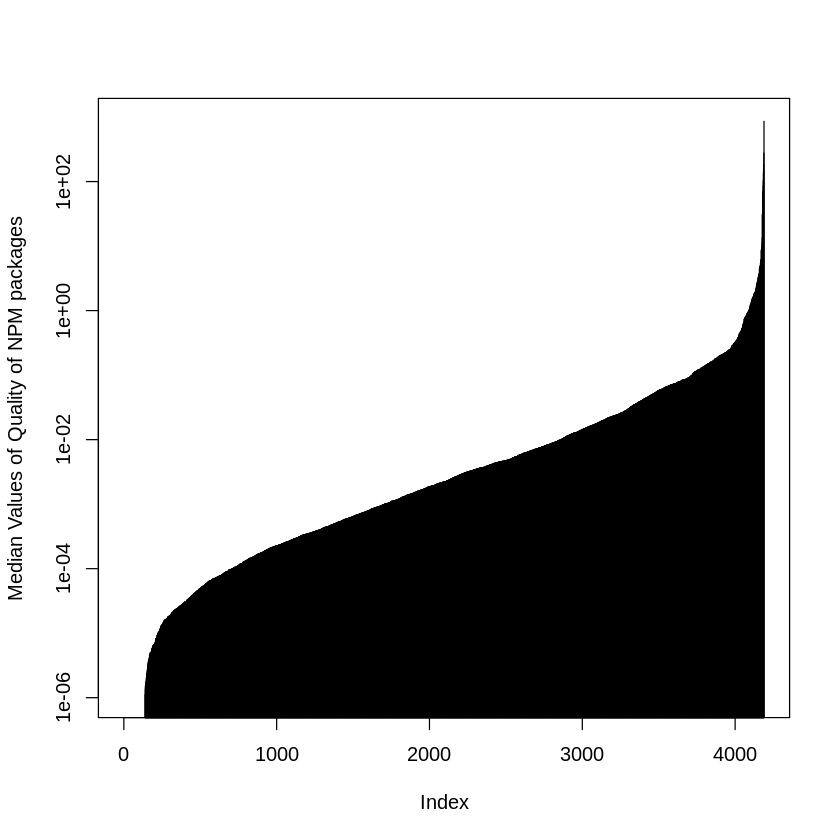}
\vspace{-10pt}
\caption{Histogram of Median Values of Quality of NPM packages}
\label{fig:qNPM}
\vspace{-10pt}
\end{figure}

To check the quality of different packages, we looked at the minimum and median value of the quality metric for the 4430 packages. We didn't consider the absolute maximum value, since some packages had zero downloads for a few days, driving the value of the quality metric to infinity. So, we used the $90^{th}$ quantile value as a proxy for the maximum value. We looked at the packages for which the value of the quality metric was more than 1. The threshold was chosen because we were looking at the packages of really high values of the quality metric, and thus were of poor quality. We found that for 340 packages the $90^{th}$ quantile value of the quality metric was more than 1, i.e. they had more than 1 issue reported against them per download. The number was 100 and 3 when we looked at the median and minimum values respectively. The three packages for which the total number of issues over the number of daily downloads was more than 1 were '@ngrx/store', '@protobufjs/fetch', and '@protobufjs/inquire'. Overall, we found that the packages for which the value of the quality metric was more than 1 were mostly packages from a big project that were relatively less downloaded, \emph{e.g.} `babel-plugin-transform-es2015-bLOCk-scoped-functions' from babel project, `react-scripts' from facebook-react project etc. There were a few other packages that had very few downloads during most of its life-cycle since 2015, but had an increase in popularity later on, and thus were selected in our list of packages. However, since they had very few downloads for a long time, the median or maximum value of the quality metric was more than 1 (\emph{e.g.} `bubleify'). 
For illustration, in Figure~\ref{fig:qNPM} we are showing the histogram of the median value of the quality metric for the 4430 NPM packages, which gives some idea about the overall quality distribution of the packages in the NPM ecosystem. We can see that around 75\% of the packages in NPM have a median value of the quality metric less than 0.01, which mean, overall, for around 75\% of the NPM packages, less than 1 in 100 regular users ever (since we are looking at the total number of issues) file an issue.

Further inspection showed that the value of the quality variable increases with time for almost half of the packages (2030 out of 4430 packages, 45.8\%), unlike what we observed for the mobile applications, where for almost all of the releases of the three softwares, the value of the quality variable decreased with time.

\begin{figure}[!t]
\begin{minipage}{.45\textwidth}
\centering
\includegraphics[width=\linewidth]{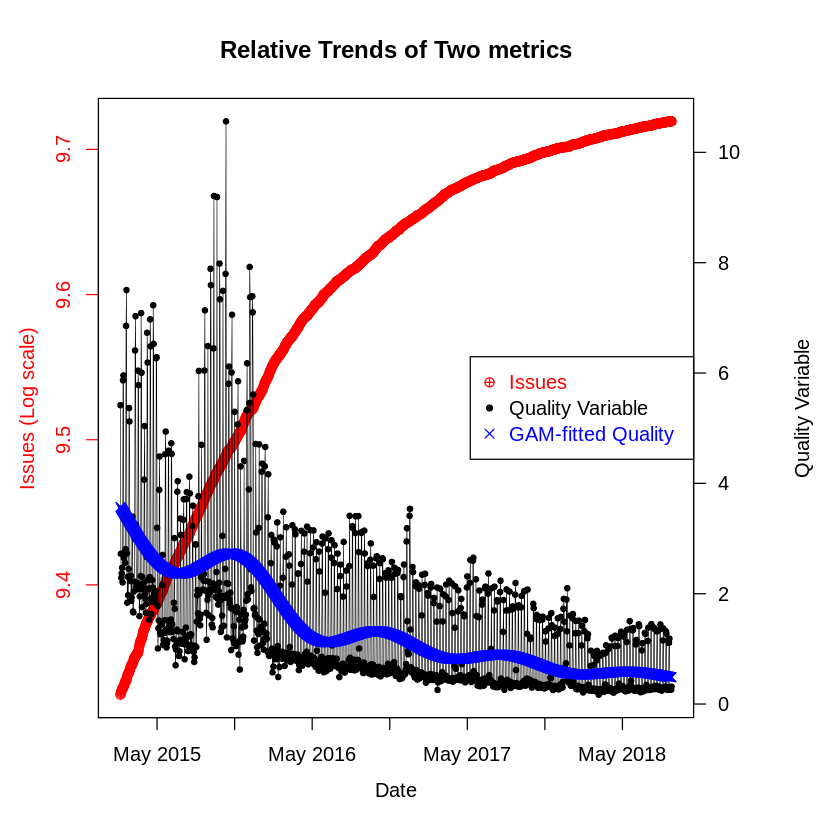}
\caption{Timeline for NPM package: angular}
\label{fig:tNa}
\end{minipage}
\hfill
\begin{minipage}{.45\textwidth}
\centering
\includegraphics[width=\linewidth]{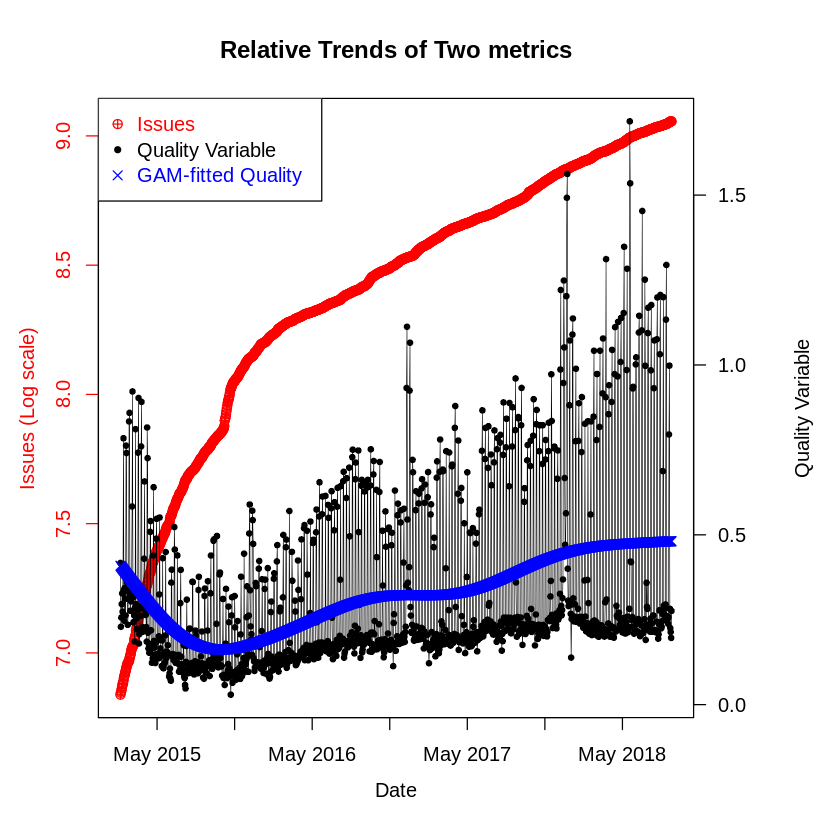}
\caption{Timeline  for NPM package: babel}
\label{fig:tNb}
\end{minipage}
\vspace{-10pt}
\end{figure}

\begin{figure}[!t]
\begin{minipage}{.45\textwidth}
\centering
\includegraphics[width=\linewidth]{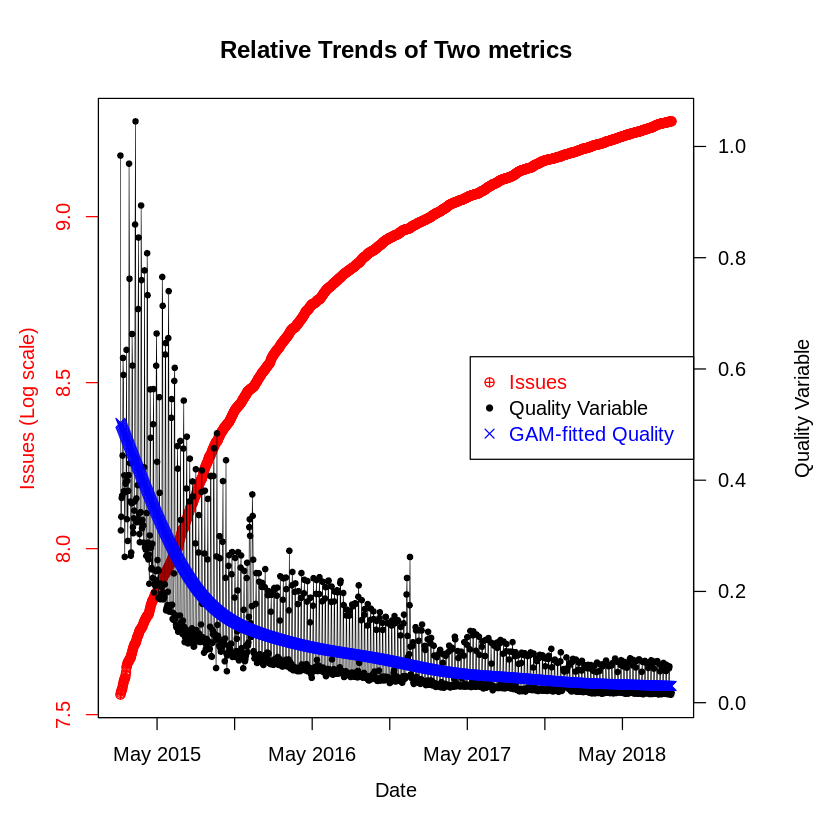}
\caption{Timeline  for NPM package: eslint}
\label{fig:tNe}
\end{minipage}
\hfill
\begin{minipage}{.45\textwidth}
\centering
\includegraphics[width=\linewidth]{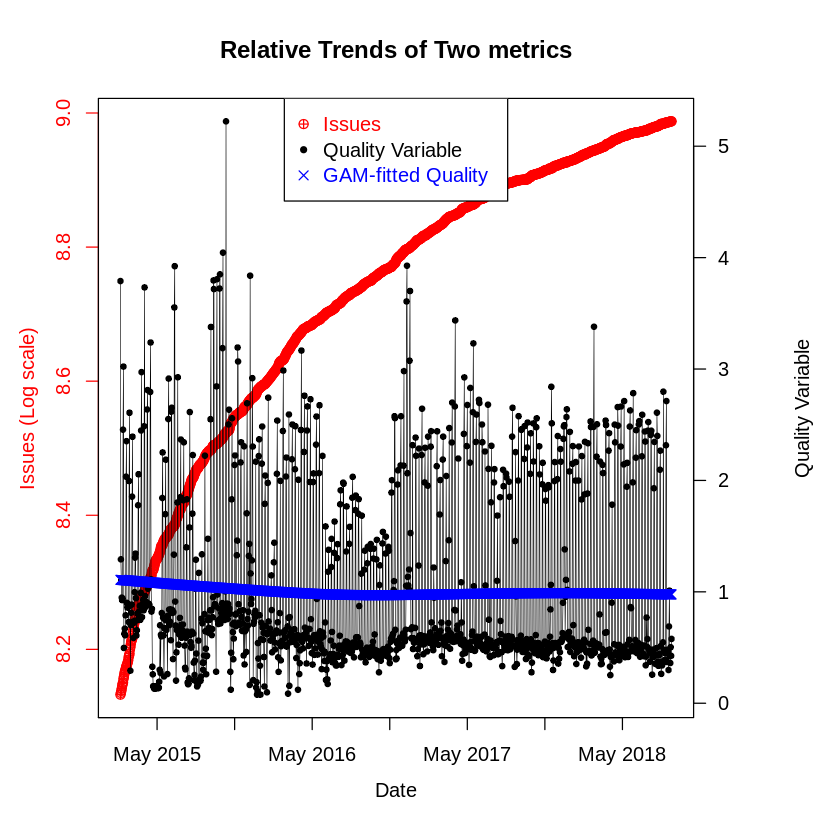}
\caption{Timeline  for NPM package: ember-cli}
\label{fig:tNel}
\end{minipage}
\vspace{-10pt}
\end{figure}

Here we also show the timelines comparing the trend of the quality variable we defined (\emph{i.e.} in this case, number of issues per download), along with a fitted line that was fitted using the Generalized Additive Model (GAM), and the number of issues, for a few selected well-known NPM packages for illustration. 
The selected NPM packages are quite popular and have a large number of issues reported against them, so so plotted the number of issues in log scale. We can see that for all the four cases, the number of issues keep increasing with a decreasing slope, but the quality measure follows different trends for the four cases.
We see that the quality measure for ``angular''(Figure~\ref{fig:tNa}) and ``eslint''(Figure~\ref{fig:tNe}) have a trend similar to what we saw for the mobile apps, with the value of the quality variable decreasing with time, but ``babel''(Figure~\ref{fig:tNb}) is showing an increase in the value, followed by an initial decrease, while for ``ember-cli'' (Figure~\ref{fig:tNel}), the trend is almost constant over time.
This result clearly show the necessity of normalizing the number of issues, which is a measure of software faults, by usage parameter like the number of downloads before using it as a measure for software quality.

\vspace{-20pt}
\section{Discussion}\label{s:implication}

Our analysis makes it evident that the number of users  is one of the most important variables 
in explaining various post-release software failure metrics, as seen in all three of the mobile 
applications as well as for the NPM packages. 
The analysis also indicates that, for the mobile applications, more new users for 
a release would mean more exceptions would be found for the software
and, for the GA releases of both the apps analyzed, longer activity for the release 
(the duration of a release measures how long a release is actively used by users,
not the time between two releases, since the releases overlap).  This
suggests that users  may be reluctant to upgrade (or are encouraged to stay) on better
-quality releases. For the NPM packages as well, a higher number of downloads 
indicated a larger number of issues. Our findings are in agreement with findings
of~\cite{IQ08,hmps15,mockus2005predictors} that consider 
post-release defects for a completely different server software system.

The release date also affects no. of exceptions for the mobile applications, as can 
be observed by looking at the coefficients. It provides some insight on
how this software has evolved.  Even after adjusting for the effect the number of
users have on the number of exceptions, the number of exceptions are
increasing with time for the Android app, whereas is decreases for the iOS app. 
This may indicate that the software, the OS, as well as the hardware could be becoming 
more complex with time, which is consistent with a rapid growth of
functionality and the size of associated code base. The Android app is seeing more crashes due to the variations in the devices and the OS, whereas for iOS, since the devices as well as the OS versions are tightly controlled, the users are seeing less issues, although we have no explicit evidence to support our speculation. 

An interesting observation from the model is the \textit{lack of any direct relationship 
between exceptions and the intensity or frequency of usage}.
One possibility is that exceptions happen for specific Android/
iOS version/ Phone combination and the way each user is exercising application's
functionality. Users for whom the application crashes must wait for
the next release. This would lead to the observed phenomena where
only the new users increase the number of crashes, which was observed more 
clearly from the timeline of crashes as well. The duration 
an application is used by individual users was found to have a much
smaller effect on reported defects than the number of
new users in prior work~\cite{hmps15,IQ08,MZL05} as well. In particular, it was observed
that most of the issues happen \textit{soon after deploying the release} 
and the chances of reporting a defect for a new release drops 
very rapidly with time after installation.

\hypobox{
We found that the exceptions are a result of more new users and the extent
of usage does not appear to have a direct effect on the number of new users.
}

Our study of the interrelationship between the code complexity metrics, 
the number of downloads, and the number of issues for the 520 NPM packages revealed that 
even after accounting for the code complexity metrics, the number of issues had strong 
dependence on the number of downloads. This result confirms the hypotheses we discussed in 
Section~\ref{s:motiv}, showing (1) The number of issues depend on both internal factors like 
code complexity, and external factors like usage, and (2) From the BN model we observed that 
the internal code complexity metrics had no influence on the number of downloads, which is an 
external factor.  

\hypobox{
We found that usage of NPM packages, measured by downloads, was a significant predictor for 
issues even after taking the code complexity metrics into consideration.
}

We found that among the code complexity metrics, the average per-function count of logical 
lines of code (\textit{loc}) was the most important predictor for modeling the number of 
issues, and the effect of the other factors was much less pronounced. From the BN model we 
found the relationship between the \textit{loc} and the number of issues to be negative, 
\textit{i.e.} the modules with more average per-function logical lines of code were seeing 
fewer issues. However, given that this measure is the per-function lines of code, it could 
indicate that simpler modules with fewer functions are less likely to have issues reported 
against them.

It is worth mentioning that the most likely reason our models for the mobile applications 
showed a relatively poor predictive performance is that we did not have any internal measures 
like the code complexity metrics in those models, and given the number of issues depend on 
internal as well as external factors (what we saw from the results of our NPM analysis), not 
having the internal factors affected the predictive performance of those models. Getting code 
complexity metrics for the closed source mobile applications proved difficult, due to the 
proprietary nature of the code, and the fact that the development teams worked on multiple 
releases at the same time further complicated getting the code complexity measures for 
particular releases. Therefore, we investigated the NPM packages, which are open-source, to 
verify the impact of usage after taking the code complexity measures into consideration.  
Since we had both the internal as well as the external factors in the models for the 520 NPM 
packages, the predictive performance was much better.

From the timeline analysis of the 4430 NPM packages, we observed that the number of downloads
 is a significant predictor for the number of issues for  most of them, and when controlled for 
the calendar date, which compensated for the variations in the downloads by automated sources, 
it was a significant predictor for all the 4430 NPM packages. So, a similar quality measure was
 used for this case as well. 
We found by looking into this metric that, overall, for around 75\% of the NPM packages, less than 1 in 100 regular users ever (since we are looking at the total number of issues) file an issue.
However, unlike the three mobile apps, where the value of our quality metric decreases with time for all releases, for the NPM packages the quality metric sometimes increases or remain relatively constant over time (around 45.8\% of the time). 

Our data, scripts, and more detailed results are available in our GitHub repository: \url{https://github.com/tapjdey/release\_qual\_model}.
 
Overall, none of the three models indicate that ``Usage.Frequency'' or ``Usage.Intensity''
 have any effect on the ``Quality'' variable. We, therefore, suggest that the exceptions per 
 user, or a metric similar to that, can be used as a software development quality metric to
 objectively compare quality of different releases. While the measure is not very novel or 
sophisticated (Post-release defect density calculated as a proportion of users who experience 
an issue within a certain period after installing or upgrading to a new release has been 
proposed by~\cite{mockus2008interval,mockus2005predictors} as a measure of software quality), 
it is an \textit{actionable} and \textit{easy to use} measure. 
A more sophisticated approach would require modeling the software failure measures (like 
exceptions or issues) as a function of software usage, and then use the residuals obtained 
after fitting the model to objectively compare the qualities of the
releases. Such approach may prove to be too complex for a
development team to apply.

The wider practical implication of this finding is twofold:
\begin{enumerate}
\item Our findings prove that due to the interdependence of usage and the observed number of software failures (like exceptions), \emph{any} quality measure (like number of defects, defect density, mean time between failures) that is dependent on any of these observed number software failures would misclassify some releases being better or worse than others unless the usage aspect is taken into account. The effect naturally would be more pronounced for softwares/releases with a large variation in usage. 
\item The results of our findings also suggest that these observed number of software failures do not depend on all aspects of usage, e.g. we found no dependence between usage intensity or frequency and number of observed exceptions. It suggests that to make a quality measure independent of the external factors like usage, we can not just normalize it by any usage measure, e.g. normalizing the number of exceptions by usage intensity or frequency would not make it independent of external factors. It is important to normalize by the right measure to be able to actually make the quality measure independent of usage. 
\end{enumerate}

\vspace{-20pt}
\section{Related Work}\label{s:relwork}

Although software quality has always been a common
topic in software engineering~\cite{boehm1976quantitative,kitchenham1996software}, most of
the studies have focused on pre-release data, primarily due to the
developers' concern about finding the appropriate balance between
the amount of testing required and the quality of software
(e.g.~\cite{rubin2016challenges,dalal1988should}). There have been a
number of works on predicting and improving the software quality as
well (e.g.~\cite{MHP13,zhang2015towards,KSAHMSU13,MW00}). Comparatively,
studies about post-deployment quality and dynamics have been less
frequent~\cite{li2011characterizing,kenny1993estimating}. However, a
number of studies have looked at the aspects of software quality
metrics, especially the quality perceived by the customers,
e.g.,~\cite{mockus2005predictors,IQ08,hmps15,rotella2011implementing,M14}. 
\cite{amreen2019methodology} described a general way to measure Software Quality and related metrics for Open Source Software ecosystems. 
A notable
non-academic work involves a study of mobile app monitoring
company's (Crittercism) data~\cite{crittercism12}. The author of the
news article found it necessary to normalize crash data by the
number of launches. Finally, an empirical investigation between
release frequency and quality on Mozilla Firefox has been
investigated in~\cite{khomh2012faster}. 

While Bayesian Networks have been used for software defect prediction 
for decades, the use of BNs for explanatory modeling in
empirical software engineering is still not common despite the
promise. A case for use of BNs was made
by Fenton et.al.~\cite{fenton1999critique,fenton2002software}, while the earliest publications
utilizing BNs we could find~\cite{HM03a} constructed search of the
structure based on the statistical significance of partial
correlations in the context of modeling delays in globally
distributed development. \cite{stamelos2003use,pendharkar2005probabilistic} considered
the application of Bayesian networks to prediction of effort, 
\cite{fenton2007predicting,neil1996predicting,okutan2014software} 
used Bayesian networks to predict defects, and \cite{pai2007empirical} 
used BN approach for an empirical analysis of faultiness of a software. 
On the other hand, Bayesian structure learning is a big domain in itself 
with a wide range of algorithms, but its use in software engineering context 
is not very common.

Hackbarth et al.~\cite{hackbarth2016improving} found the need to adjust defect counts in their proposed measure of software quality as perceived by customers. We propose a somewhat different measure of quality based on the number of exceptions per user. In general, software quality is a widely researched topic~\cite{kan2002metrics,kitchenham1996software,schulmeyer1992handbook} etc., but in our knowledge, this is the first model-based attempt to obtain a usage independent measure of software quality and the first attempt to model exceptions in mobile applications.

The NPM ecosystem is one of the most active and dynamic JavaScript ecosystems and~\cite{wittern2016look} presents its dependency structure and package popularity.~\cite{zerouali2018empirical} studies the dependency, specifically the lag in updating dependencies in various NPM packages while~\cite{abdalkareem2017developers} looked into the use of trivial packages as part of package dependencies for different NPM packages. \cite{dey2018software} investigated the factors affecting NPM package popularity, and~\cite{dey2019patterns} investigated the participation patterns of issue and patch creators.

The advancements proposed in this paper over the published work are focused 
on two primary areas: (1) study of the relationship between software faults (issues for NPM packages) and usage using \textbf{post-release} data in the context of two proprietary mobile applications and 4430 popular NPM packages, and (2) proposing a usage independent exception-based software quality metric based on our models.

\vspace{-10pt}
\subsection{Comparison with published results}
In this subsection we compare our findings with already reported results that studied other commercial applications. The goal of this subsection is not replicating the earlier studies, but just 
comparing the findings of our study and those of some earlier studies. We add this section to address 
the limitation of our dataset having a relatively small sample of data. 

Unfortunately, there aren't a lot of studies that looked into the interrelationship between software usage and software faults (defects or crashes).

The number of users for most of the releases we studied are very small, with a 
median of 7 users per release, although a few releases have more than 16,000 users.
On slide 22 of of his presentation~\cite{caper}, Caper Jones reported that the
number of defects increase 2 to 3 times for a 10 fold increase in the number of users
(from 1 to 10 and 10 to 100) for a software of similar complexity (between 10,000
and 100,000 function points). However, they were looking at the number of 
defects, and typically the number of 
exceptions is larger than the number of defects, because one defect could cause crashes for multiple users (or multiple crashes for a single user). The study published in~\cite{IQ08} was done for a system with many 
more users (around 4,000 to 16,000),however, they reported that for a two-fold increase in the 
number of users the number of Modification Requests (MR tickets) increase around $1.25$ times, 
which is more than what would have been predicted by our model ($1.02 - 1.04$) for Android apps, but less than what we have ($1.6$) for the iOS app. 

Although we were unable to do a direct comparison to another mobile application, due to different studies looking at different measures, these findings add more context to our result, and indicates the necessity of further studies that publish their datasets to understand the usage-fault relationship in a wider range of applications.

\vspace{-10pt}
\section{Limitations}\label{s:limitation}

The accuracy of our result is very much dependent on the Google
Analytics data. While we do not have reasons to doubt the accuracy of
the counts in Google
Analytics data, we would have liked to have better definitions of
how it determines ``New User'', ``Visit'', and, especially,
nontrivial to aggregate quantities such as ``Visits per User.'' Also,
it is not clear if Google
Analytics distorts data in any way (e.g., by applying differential 
privacy transformations) for low counts in order to protect
the privacy of the users. We do not believe it does, but we have not
conducted an experiment to validate that. 

Furthermore, the mobile applications under consideration were relatively new and
it was the first attempt for the team to deploy mobile software. As
such, much was not well documented and was rapidly
evolving over time. As mentioned earlier, we did not have the official release
dates for all releases, so we put the start date of the release as
the date on which the first usage was reported. However, we did
verify the official dates with this reported date for the releases
for which we found the release date, and they were very close, but
not always exactly the same. This should not affect
the overall result, given the total time scale of more than two
years. The release end dates, by their nature, have to be estimated 
based on user activity, since there is no way to force end user
to upgrade Android app. For recent releases, therefore, the end date 
may be censored by our data
collection date, hence the duration for these releases might be
underestimated. 

Another limitation associated with using these commercial closed-source 
mobile applications is that we had no control over the release cycle or the
variables being measured by Google Analytics. This limited our options for 
doing the analysis, sometimes 
severely. We had very few releases for the iOS application, and even the largest
dataset of GA releases of the Android application had only 173 releases. We had 
a limited number of observed variables as well. However, we were unable to obtain
any more data on the applications, forcing us to work with the limited data. 
However, we tried to increase the validity of our study by looking into three 
sets of releases for two applications, and used three different modeling approaches 
to study these datasets. The fact that we saw a strong relationship between the number
of users and exceptions in all cases has led us to have confidence in the validity
of our finding. 

It may be possible to collect numerous additional variables that may
have an impact on exceptions, for example, the number of changes to
the source code made for a release as was done
in~\cite{IQ08}. Unfortunately, due to the nature of parallel
development for multiple releases and products noted in
subsection~\ref{sub:soft}, it was virtually impossible to separate
the changes that would only affect a specific release on the Android/iOS
platform. To further complicate the matter, the mobile applications we studied
were commercial in nature, and the source code for these were not available.

Our study of the mobile applications focused on a single set of mobile applications 
from a specific domain, implemented via
a rather complex codebase and is certainly not representative of
most mobile applications that tend to be much simpler. Furthermore,
mobile applications  
may not represent other types of software further limiting external
validity of the results. However, some aspects that we see in the
specific application, such as increasing number of faults with the
number of users, has been observed in rather different contexts of
large-scale server software. This suggests that the model derived in
the study may generalize to other domains as well.

In terms of modeling aspects, there are some limitations related to the different approaches. 
The RF model was used for 10 times 2 fold cross-validation, and exhibited a rather high value 
of standard deviation in the $R^2$ value,  likely due to the small sample size.

While creating the BN model we did not cover all possible ways BNs can be applied to gain 
insight into the system. For example, we did not investigate the possible existence of any 
hidden node, or make an effort to formally establish the causal relationship between the nodes. 
We also did not investigate how the properties of one release affect the subsequent releases, 
nor did we investigate the presence of any feedback loops. Although we used the best methods 
identified from the simulation study, we did not employ any measures to verify the 
existence/non-existence of any link that appeared in the averaged bootstrapped model. 

In the simulation study, although we covered an extensive set of options, we did not try every possible combination of options for the BN structure search exercise.

We also did not use Markov Random Field analysis, which is another probabilistic graphical modeling approach. The primary reason behind choosing the BN approach was that we found an example where this method was used to successfully recover the underlying network~\cite{bnppt}. 
Moreover, it is possible to interpret a BN model as a causal model, and although we did not use that interpretation in this study, our goal is to eventually establish a causal mechanism of how usage affects the number of exceptions/defects experienced by users, so we wanted to used BN from the start.

Regarding external validity, we analyzed 520 most popular NPM packages, which
is less than 0.1\% of the total packages in the NPM ecosystem. Even during the
timeline study, we only looked at 4430 packages. These packages, however, represent
the tiny part of the NPM ecosystem that is widely used, so they constitute 
a suitable subset for our study. 

Although the study of the NPM packages had measures related to code complexity 
and usage, we didn't look into a lot of other possible variables that could 
affect the number of issues,\emph{e.g.} the number of 
dependents a package has. Although some of the issues could come 
from users of a dependent package, we didn't actively check the
origins of the issues to verify that. We also didn't look at the
releases of the packages, because of reasons mentioned before.
We didn't differentiate between the types of the issues, because we 
just wanted to see how many times a user decided to file an issue.
Overall, this study was not a direct extension of the previous work,
rather, it was an extension of the concept and its application in a different domain.

Another approach that could have been taken to make this study more similar to
the study of the mobile applications would require us to check whether or not
an issue filed for a package has a crash report. However, such an approach would 
come with different caveats, e.g. a crash could result from the limitation of the 
package, but it could also result from some bug or compatibility issue
in the web browser, or even the OS. Due to these limitations, we did not investigate 
this in this study, though, it is an interesting question we would like to address in future.

\vspace{-20pt}
\section{Conclusion}\label{s:conclusion}

From the practical perspective we have established
that an external factor like the extent of use has very 
strong relationship with the observed number
of exceptions for three large mobile applications from 
the telecommunication domain. The study of the 520 NPM packages
revealed that the effect remains noticeable even after taking the 
code complexity measures into account.
Counting exceptions, or using any other quality 
measure dependent on an observed number exceptions, or any other 
software failure metric (like number of defects or issues), 
therefore, will not accurately measure the quality of software
development process but, instead, it would strongly depend on the
extent of use. In order to produce a measure that the development
team can use to understand and improve quality of their software
development process, we proposed to normalize the observed exceptions by usage, 
specifically by number of users or any related measure if it is not 
available. Notably, a similar
normalization was previously proposed in the context of post-release
defects that also exhibited strong positive correlation with the
number of users.  As a larger proportion of applications are mobile
and/or delivered as a service, the amount of usage can be relatively
easily collected.  Consequently, not adjusting software
development measures for usage should not be considered as an
excusable practice.

From theoretical perspective, we provided the explanation of the
relationships among post-deployment quantities using Bayesian Networks, which
allow for exploration of relationships among all variables and
empirical determination of the relationships exhibited in a
particular dataset.  For all three mobile softwares analyzed, the number of users was found to be the most significant predictor with both the models. It would be preferable to have each release as a separate
categorical predictor, but because for simplicity we chose to use
only one observation per release.

We also established
that it is possible to predict exceptions using Random Forest
modeling techniques and that usage plays a key role for the accuracy
of these predictions. However, the performance of the predictive model 
was not consistently good, since we did not have any internal factor as a 
predictor and, as noted above, prediction is a different task
than explanation and, even though it often yields more accurate
results, the prediction results may be harder to explain to
developers or managers and, therefore, harder to act upon. We
believe the findings do have a message for the voluminous research
in defect prediction. While defects are not exceptions, usage was
also found to affect post-release defects in a similar
manner~\cite{caper,hmps15,mockus2005predictors}. It would, therefore, be advisable to 
incorporate forecasts of usage into defect prediction models  to increase their accuracy.

Our analysis of the NPM packages established that our approach is extendable to other domains as well. 
The study revealed that even a less accurate measure of usage like downloads, which, for NPM packages, 
is a mix of downloads by human users and automated sources, is an important predictor for the number 
of issues reported, which again is a weakly similar measure to the number of crashes or bugs. 
So, our approach can be applied to any situation similar to the ones we studied, even when only 
proxy measures for usage and crashes/ bugs are available. The study also revealed the 
importance 
of taking software usage into account even in the presence of code complexity measures.

We hope that this work will spur more research on software engineering
aspects in post-deployment stage because, like mobile applications, 
modern web applications are even more reliant on usage monitoring not
simply from the perspective of crash counting but also because the
usability or even revenue stream from the software applications
critically depends on how users behave. 

From the practical perspective, we hope that any mobile or web
software project can easily apply and refine the presented approach
of using Google Analytics data to improve the quality of their
software.  Any Android OS or Apple iOS mobile application can use
Google Analytics to monitor application usage and crashes, so the
approach should be widely applicable. Despite that, we 
are not aware of any prior empirical study that would leverages Google
Analytics or similar data for software quality modeling.

The result of our simulation study should also be useful for practitioners 
using Bayesian Network structure search techniques for choosing the best performing methods.

Finally, much more work is needed to gather additional empirical
evidence of how software behaves post-deployment. It is important to
note that Google Analytics data is available only for application
developers, so while each project has the ability to see their app's
performance, they can not see data for software created by other
organizations. This can be addressed by a) projects sharing theirs
post-deployment data (we have not seen examples of that); or b)
publishing findings based on such data in cases such as ours,
where the data itself would be impossible to release publicly since
it involves numerous, often enterprise, customers who may not agree.

\begin{acknowledgements}
This work was supported by the National Science Foundation (U.S.) under Grant No. 1633437 and Grant No. 1901102.
\end{acknowledgements}

\bibliographystyle{spmpsci}      
\bibliography{reference.bib}   

\begin{thebibliography}{10}
\providecommand{\url}[1]{{#1}}
\providecommand{\urlprefix}{URL }
\expandafter\ifx\csname urlstyle\endcsname\relax
  \providecommand{\doi}[1]{DOI~\discretionary{}{}{}#1}\else
  \providecommand{\doi}{DOI~\discretionary{}{}{}\begingroup
  \urlstyle{rm}\Url}\fi

\bibitem{abdalkareem2017developers}
Abdalkareem, R., Nourry, O., Wehaibi, S., Mujahid, S., Shihab, E.: Why do
  developers use trivial packages? an empirical case study on npm.
\newblock In: Proceedings of the 2017 11th Joint Meeting on Foundations of
  Software Engineering, pp. 385--395. ACM (2017)

\bibitem{pcalgR}
{Alain Hauser, Peter Buehlmann}: Characterization and greedy learning of
  interventional markov equivalence classes of directed acyclic graphs.
\newblock Journal of Machine Learning Research, \textbf{13}, 2409--2464 (2012).
\newblock \urlprefix\url{http://jmlr.org/papers/v13/hauser12a.html}

\bibitem{amreen2019methodology}
Amreen, S., Bichescu, B., Bradley, R., Dey, T., Ma, Y., Mockus, A., Mousavi,
  S., Zaretzki, R.: A methodology for measuring floss ecosystems.
\newblock In: Towards Engineering Free/Libre Open Source Software (FLOSS)
  Ecosystems for Impact and Sustainability, pp. 1--29. Springer, Singapore
  (2019)

\bibitem{catnetR}
Balov, N., Salzman, P.: catnet: Categorical Bayesian Network Inference (2016).
\newblock \urlprefix\url{https://CRAN.R-project.org/package=catnet}.
\newblock R package version 1.15.0

\bibitem{boehm1976quantitative}
Boehm, B.W., Brown, J.R., Lipow, M.: Quantitative evaluation of software
  quality.
\newblock In: Proceedings of the 2nd international conference on Software
  engineering, pp. 592--605. IEEE Computer Society Press (1976)

\bibitem{borges2016understanding}
Borges, H., Hora, A., Valente, M.T.: Understanding the factors that impact the
  popularity of github repositories.
\newblock In: Software Maintenance and Evolution (ICSME), 2016 IEEE
  International Conference on, pp. 334--344. IEEE (2016)

\bibitem{dealR}
Bottcher, S.G., Dethlefsen., C.: deal: Learning Bayesian Networks with Mixed
  Variables (2013).
\newblock \urlprefix\url{https://CRAN.R-project.org/package=deal}.
\newblock R package version 1.2-37

\bibitem{briand2000exploring}
Briand, L.C., W{\"u}st, J., Daly, J.W., Porter, D.V.: Exploring the
  relationships between design measures and software quality in object-oriented
  systems.
\newblock Journal of systems and software \textbf{51}(3), 245--273 (2000)

\bibitem{chatzidimitriou2018npm}
Chatzidimitriou, K.C., Papamichail, M.D., Diamantopoulos, T., Tsapanos, M.,
  Symeonidis, A.L.: npm-miner: An infrastructure for measuring the quality of
  the npm registry.
\newblock In: Proceedings of the 15th International Conference on Mining
  Software Repositories, pp. 42--45. ACM (2018)

\bibitem{chickering1996learning}
Chickering, D.M.: Learning bayesian networks is np-complete.
\newblock Learning from data: Artificial intelligence and statistics V
  \textbf{112}, 121--130 (1996)

\bibitem{chlebus1998finding}
Chlebus, B.S., Nguyen, S.H.: On finding optimal discretizations for two
  attributes.
\newblock In: International Conference on Rough Sets and Current Trends in
  Computing, pp. 537--544. Springer (1998)

\bibitem{dalal1988should}
Dalal, S.R., Mallows, C.L.: When should one stop testing software?
\newblock Journal of the American Statistical Association \textbf{83}(403),
  872--879 (1988)

\bibitem{npmpkg}
David:  (2014).
\newblock \urlprefix\url{https://developers.slashdot.org/story/17/01/14/
  0222245/nodejss-npm-is-now-the-largest-package-registry-in-the-world}

\bibitem{dey2019patterns}
Dey, T., Ma, Y., Mockus, A.: Patterns of effort contribution and demand and
  user classification based on participation patterns in npm ecosystem.
\newblock arXiv preprint arXiv:1907.06538  (2019)

\bibitem{dey2018software}
Dey, T., Mockus, A.: Are software dependency supply chain metrics useful in
  predicting change of popularity of npm packages?
\newblock In: Proceedings of the 14th International Conference on Predictive
  Models and Data Analytics in Software Engineering, pp. 66--69. ACM (2018)

\bibitem{dey2018modeling}
Dey, T., Mockus, A.: Modeling relationship between post-release faults and
  usage in mobile software.
\newblock In: Proceedings of the 14th International Conference on Predictive
  Models and Data Analytics in Software Engineering, pp. 56--65. ACM (2018)

\bibitem{amhp14}
Duc, A.N., Mockus, A., Hackbarth, R., Palframan, J.: Forking and coordination
  in multi-platform development: a case study.
\newblock In: ESEM, pp. 59:1--59:10. Torino, Italy (2014).
\newblock \urlprefix\url{http://dl.acm.org/authorize?N14215}

\bibitem{fenton2002software}
Fenton, N., Krause, P., Neil, M.: Software measurement: Uncertainty and causal
  modeling.
\newblock IEEE software \textbf{19}(4), 116--122 (2002)

\bibitem{fenton2008using}
Fenton, N., Neil, M., Marquez, D.: Using bayesian networks to predict software
  defects and reliability.
\newblock Proceedings of the Institution of Mechanical Engineers, Part O:
  Journal of Risk and Reliability \textbf{222}(4), 701--712 (2008)

\bibitem{fenton2007predicting}
Fenton, N., Neil, M., Marsh, W., Hearty, P., Marquez, D., Krause, P., Mishra,
  R.: Predicting software defects in varying development lifecycles using
  bayesian nets.
\newblock Information and Software Technology \textbf{49}(1), 32--43 (2007)

\bibitem{fenton1999critique}
Fenton, N.E., Neil, M.: A critique of software defect prediction models.
\newblock IEEE Transactions on software engineering \textbf{25}(5), 675--689
  (1999)

\bibitem{friedman1999data}
Friedman, N., Goldszmidt, M., Wyner, A.: Data analysis with bayesian networks:
  A bootstrap approach.
\newblock In: Proceedings of the Fifteenth conference on Uncertainty in
  artificial intelligence, pp. 196--205. Morgan Kaufmann Publishers Inc. (1999)

\bibitem{crittercism12}
Geron, T.: Do ios apps crash more than android apps? a data dive (2012).
\newblock
  \url{https://www.forbes.com/sites/tomiogeron/2012/02/02/does-ios-crash-more-than-android-a-data-dive}

\bibitem{hmps15}
Hackbarth, R., Mockus, A., Palframan, J., Sethi, R.: Customer quality
  improvement of software systems.
\newblock Software, IEEE \textbf{33}(4), 40--45 (2016).
\newblock \urlprefix\url{papers/cqm2.pdf}

\bibitem{hackbarth2016improving}
Hackbarth, R., Mockus, A., Palframan, J., Sethi, R.: Improving software quality
  as customers perceive it.
\newblock IEEE Software \textbf{33}(4), 40--45 (2016)

\bibitem{arulesR}
Hahsler, M., Chelluboina, S., Hornik, K., Buchta, C.: The arules r-package
  ecosystem: Analyzing interesting patterns from large transaction datasets.
\newblock Journal of Machine Learning Research \textbf{12}, 1977--1981 (2011).
\newblock \urlprefix\url{http://jmlr.csail.mit.edu/papers/v12/hahsler11a.html}

\bibitem{hartemink2001principled}
Hartemink, A.J.: Principled computational methods for the validation and
  discovery of genetic regulatory networks.
\newblock Ph.D. thesis, Massachusetts Institute of Technology (2001)

\bibitem{HM03a}
Herbsleb, J.D., Mockus, A.: An empirical study of speed and communication in
  globally-distributed software development.
\newblock IEEE Transactions on Software Engineering \textbf{29}(6), 481--494
  (2003).
\newblock \urlprefix\url{papers/delay.pdf}

\bibitem{caper}
Jones, C.: Software quality in 2011: A survey of the state of the art.
\newblock \url{http://sqgne.org/presentations/2011-12/Jones-Sep-2011.pdf}
  (2011).
\newblock President, Namcook Analytics LLC, www.Namcook.com Email:
  Capers.Jones3@GMAILcom

\bibitem{pcalgR2}
Kalisch, M., M\"achler, M., Colombo, D., Maathuis, M.H., B\"uhlmann, P.: Causal
  inference using graphical models with the {R} package {pcalg}.
\newblock Journal of Statistical Software \textbf{47}(11), 1--26 (2012).
\newblock \urlprefix\url{http://www.jstatsoft.org/v47/i11/}

\bibitem{KSAHMSU13}
Kamei, Y., Shihab, E., Adams, B., Hassan, A.E., Mockus, A., Sinha, A.,
  Ubayashi, N.: A large-scale empirical study of just-in-time quality
  assurance.
\newblock IEEE Transactions on Software Engineering \textbf{39}(6), 757--773
  (2013).
\newblock
  \urlprefix\url{http://doi.ieeecomputersociety.org/10.1109/TSE.2012.70}

\bibitem{kan2002metrics}
Kan, S.H.: Metrics and models in software quality engineering.
\newblock Addison-Wesley Longman Publishing Co., Inc. (2002)

\bibitem{kenny1993estimating}
Kenny, G.Q.: Estimating defects in commercial software during operational use.
\newblock IEEE Transactions on Reliability \textbf{42}(1), 107--115 (1993)

\bibitem{khomh2012faster}
Khomh, F., Dhaliwal, T., Zou, Y., Adams, B.: Do faster releases improve
  software quality?: an empirical case study of mozilla firefox.
\newblock In: Proceedings of the 9th IEEE Working Conference on Mining Software
  Repositories, pp. 179--188. IEEE Press (2012)

\bibitem{kitchenham1996software}
Kitchenham, B., Pfleeger, S.L.: Software quality: the elusive target [special
  issues section].
\newblock IEEE software \textbf{13}(1), 12--21 (1996)

\bibitem{koller2009probabilistic}
Koller, D., Friedman, N.: Probabilistic graphical models: principles and
  techniques.
\newblock MIT press (2009)

\bibitem{kononenko2015investigating}
Kononenko, O., Baysal, O., Guerrouj, L., Cao, Y., Godfrey, M.W.: Investigating
  code review quality: Do people and participation matter?
\newblock In: Software Maintenance and Evolution (ICSME), 2015 IEEE
  International Conference on, pp. 111--120. IEEE (2015)

\bibitem{li2011characterizing}
Li, P.L., Kivett, R., Zhan, Z., Jeon, S.e., Nagappan, N., Murphy, B., Ko, A.J.:
  Characterizing the differences between pre-and post-release versions of
  software.
\newblock In: Proceedings of the 33rd International Conference on Software
  Engineering, pp. 716--725. ACM (2011)

\bibitem{bnlearnR}
{Marco Scutari}: Learning bayesian networks with the bnlearn r package.
\newblock Journal of Statistical Software \textbf{35}(3), 1--22 (2010).
\newblock \urlprefix\url{http://www.jstatsoft.org/v35/i03/}

\bibitem{mcintosh2014impact}
McIntosh, S., Kamei, Y., Adams, B., Hassan, A.E.: The impact of code review
  coverage and code review participation on software quality: A case study of
  the qt, vtk, and itk projects.
\newblock In: Proceedings of the 11th Working Conference on Mining Software
  Repositories, pp. 192--201. ACM (2014)

\bibitem{mcintosh2015emse}
Mcintosh, S., Kamei, Y., Adams, B., Hassan, A.E.: An empirical study of the
  impact of modern code review practices on software quality.
\newblock Empirical Softw. Engg. \textbf{21}(5), 2146--2189 (2016).
\newblock \doi{10.1007/s10664-015-9381-9}.
\newblock \urlprefix\url{http://dx.doi.org/10.1007/s10664-015-9381-9}

\bibitem{Changes07}
Mockus, A.: Software support tools and experimental work.
\newblock In: V.~Basili, et~al (eds.) Empirical Software Engineering Issues:
  Critical Assessments and Future Directions, vol. LNCS 4336, pp. 91--99.
  Springer (2007).
\newblock \urlprefix\url{papers/SSTaEW.pdf}

\bibitem{mockuskeynote}
Mockus, A.: Law of minor release: More bugs implies better software quality.
\newblock http://mockus.org/papers/IWPSE13.pdf (2013).
\newblock International Workshop on Principles of Software Evolution, St
  Petersburg, Russia, Aug 18-19 2013. Keynote

\bibitem{M14}
Mockus, A.: Engineering big data solutions.
\newblock In: ICSE'14 FOSE, pp. 85--99 (2014).
\newblock \urlprefix\url{http://dl.acm.org/authorize?N14216}

\bibitem{MHP13}
Mockus, A., Hackbarth, R., Palframan, J.: Risky files: An approach to focus
  quality improvement effort.
\newblock In: 9th Joint Meeting of the European Software Engineering Conference
  and the ACM SIGSOFT Symposium on the Foundations of Software Engineering, pp.
  691--694 (2013).
\newblock \urlprefix\url{http://dl.acm.org/authorize?6845890}

\bibitem{IQ08}
Mockus, A., Weiss, D.: Interval quality: Relating customer-perceived quality to
  process quality.
\newblock In: 2008 International Conference on Software Engineering, pp.
  733--740. ACM Press, Leipzig, Germany (2008).
\newblock \urlprefix\url{http://dl.acm.org/authorize?063910}

\bibitem{mockus2008interval}
Mockus, A., Weiss, D.: Interval quality: Relating customer-perceived quality to
  process quality.
\newblock In: Proceedings of the 30th international conference on Software
  engineering, pp. 723--732. ACM (2008)

\bibitem{MW00}
Mockus, A., Weiss, D.M.: Predicting risk of software changes.
\newblock Bell Labs Technical Journal \textbf{5}(2), 169--180 (2000).
\newblock \urlprefix\url{papers/bltj13.pdf}

\bibitem{MZL05}
Mockus, A., Zhang, P., Li, P.: Drivers for customer perceived software quality.
\newblock In: ICSE 2005, pp. 225--233. ACM Press, St Louis, Missouri (2005).
\newblock \urlprefix\url{http://dl.acm.org/authorize?860140}

\bibitem{mockus2005predictors}
Mockus, A., Zhang, P., Li, P.L.: Predictors of customer perceived software
  quality.
\newblock In: Software Engineering, 2005. ICSE 2005. Proceedings. 27th
  International Conference on, pp. 225--233. IEEE (2005)

\bibitem{negRsq}
(https://stats.stackexchange.com/users/25/harvey motulsky), H.M.: When is r
  squared negative?
\newblock Cross Validated.
\newblock \urlprefix\url{https://stats.stackexchange.com/q/12991}.
\newblock URL:https://stats.stackexchange.com/q/12991 (version: 2014-05-06)

\bibitem{nagarajan2013bayesian}
Nagarajan, R., Scutari, M., L{\`e}bre, S.: Bayesian networks in r.
\newblock Springer \textbf{122}, 125--127 (2013)

\bibitem{neil1996predicting}
Neil, M., Fenton, N.: Predicting software quality using bayesian belief
  networks.
\newblock In: Proceedings of the 21st Annual Software Engineering Workshop, pp.
  217--230. NASA Goddard Space Flight Centre (1996)

\bibitem{okutan2014software}
Okutan, A., Y{\i}ld{\i}z, O.T.: Software defect prediction using bayesian
  networks.
\newblock Empirical Software Engineering \textbf{19}(1), 154--181 (2014)

\bibitem{pai2007empirical}
Pai, G.J., Dugan, J.B.: Empirical analysis of software fault content and fault
  proneness using bayesian methods.
\newblock IEEE Transactions on software Engineering \textbf{33}(10), 675--686
  (2007)

\bibitem{pearl2011bayesian}
Pearl, J.: Bayesian networks.
\newblock Department of Statistics, UCLA  (2011)

\bibitem{pendharkar2005probabilistic}
Pendharkar, P.C., Subramanian, G.H., Rodger, J.A.: A probabilistic model for
  predicting software development effort.
\newblock IEEE Transactions on software engineering \textbf{31}(7), 615--624
  (2005)

\bibitem{perez2006supervised}
Perez, A., Larranaga, P., Inza, I.: Supervised classification with conditional
  gaussian networks: Increasing the structure complexity from naive bayes.
\newblock International Journal of Approximate Reasoning \textbf{43}(1), 1--25
  (2006)

\bibitem{R}
{R Core Team}: R: A Language and Environment for Statistical Computing.
\newblock R Foundation for Statistical Computing, Vienna, Austria (2017).
\newblock \urlprefix\url{https://www.R-project.org/}

\bibitem{rigby2013convergent}
Rigby, P.C., Bird, C.: Convergent contemporary software peer review practices.
\newblock In: Proceedings of the 2013 9th Joint Meeting on Foundations of
  Software Engineering, pp. 202--212. ACM (2013)

\bibitem{rotella2011implementing}
Rotella, P., Chulani, S.: Implementing quality metrics and goals at the
  corporate level.
\newblock In: Proceedings of the 8th Working Conference on Mining Software
  Repositories, pp. 113--122. ACM (2011)

\bibitem{rubin2016challenges}
Rubin, J., Rinard, M.: The challenges of staying together while moving fast: An
  exploratory study.
\newblock In: Proceedings of the 38th International Conference on Software
  Engineering, pp. 982--993. ACM (2016)

\bibitem{schulmeyer1992handbook}
Schulmeyer, G.G., McManus, J.I.: Handbook of software quality assurance.
\newblock Van Nostrand Reinhold Co. (1992)

\bibitem{bnppt}
Scutari, M.: Learning bayesian networks in r, an example in systems biology
  (2013).
\newblock \url{http://www.bnlearn.com/about/slides/slides-useRconf13.pdf}

\bibitem{scutari2010introduction}
Scutari, M., Strimmer, K.: Introduction to graphical modelling.
\newblock arXiv preprint arXiv:1005.1036  (2010)

\bibitem{shmueli2010explain}
Shmueli, G.: To explain or to predict?
\newblock Statistical science pp. 289--310 (2010)

\bibitem{sober2002instrumentalism}
Sober, E.: Instrumentalism, parsimony, and the akaike framework.
\newblock Philosophy of Science \textbf{69}(S3), S112--S123 (2002)

\bibitem{stamelos2003use}
Stamelos, I., Angelis, L., Dimou, P., Sakellaris, E.: On the use of bayesian
  belief networks for the prediction of software productivity.
\newblock Information and Software Technology \textbf{45}(1), 51--60 (2003)

\bibitem{subramanyam2003empirical}
Subramanyam, R., Krishnan, M.S.: Empirical analysis of ck metrics for
  object-oriented design complexity: Implications for software defects.
\newblock IEEE Transactions on software engineering \textbf{29}(4), 297--310
  (2003)

\bibitem{npmdl}
Voss, L.: numeric precision matters: how npm download counts work (2014).
\newblock
  \urlprefix\url{https://blog.npmjs.org/post/92574016600/numeric-precision-matters-how-npm-download-counts}

\bibitem{npmpop}
Voss, L.: The state of javascript frameworks, 2017 (2018).
\newblock
  \urlprefix\url{https://www.npmjs.com/npm/state-of-javascript-frameworks-2017-part-1}

\bibitem{wittern2016look}
Wittern, E., Suter, P., Rajagopalan, S.: A look at the dynamics of the
  javascript package ecosystem.
\newblock In: Mining Software Repositories (MSR), 2016 IEEE/ACM 13th Working
  Conference on, pp. 351--361. IEEE (2016)

\bibitem{yu2002predicting}
Yu, P., Systa, T., Muller, H.: Predicting fault-proneness using oo metrics. an
  industrial case study.
\newblock In: Software Maintenance and Reengineering, 2002. Proceedings. Sixth
  European Conference on, pp. 99--107. IEEE (2002)

\bibitem{zerouali2018empirical}
Zerouali, A., Constantinou, E., Mens, T., Robles, G., Gonz{\'a}lez-Barahona,
  J.: An empirical analysis of technical lag in npm package dependencies.
\newblock In: International Conference on Software Reuse, pp. 95--110. Springer
  (2018)

\bibitem{zhang2015towards}
Zhang, F., Mockus, A., Keivanloo, I., Zou, Y.: Towards building a universal
  defect prediction model with rank transformed predictors.
\newblock Empirical Software Engineering pp. 1--39 (2015)

\bibitem{zmz15}
Zheng, Q., Mockus, A., Zhou, M.: A method to identify and correct problematic
  software activity data: Exploiting capacity constraints and data
  redundancies.
\newblock In: {ESEC/FSE'15}, pp. 637--648. ACM, Bergamo, Italy (2015).
\newblock \urlprefix\url{http://dl.acm.org/authorize?N14200}

\end{thebibliography}

\end{document}